\documentclass[twocolumn]{aastex63}
\usepackage{amsmath}

\newcommand{\paperi}{Paper I}

\newcommand{\Nres}{N_{\rm res}}
\newcommand{\Lbox}{L_{\rm box}}
\newcommand{\delx}{\Delta x}

\newcommand{\mdot}{\dot{M}_w}
\newcommand{\pdot}{\dot{p}_w}
\newcommand{\Lwind}{\mathcal{L}_w}
\newcommand{\Qo}{Q_0}

\newcommand{\RWeavert}{R_{{\rm ED},\theta}} 

\newcommand{\prWeavert}{p_{{\rm ED},\theta}}

\newcommand{\RMDa}{R_{{\rm MD},\alpha}}
\newcommand{\prMDa}{p_{{\rm MD},\alpha}}

\newcommand{\Req}{R_{\rm eq}}
\newcommand{\teq}{t_{\rm eq}}

\newcommand{\RSt}{R_{\rm St}}
\newcommand{\RSp}{R_{\rm Sp}}

\newcommand{\tdio}{t_{{\rm d},i,0}}

\newcommand{\trec}{t_{\rm rec}}
\newcommand{\ReqRSt}{\zeta}

\newcommand{\nhat}{\mathbf{\hat{n}}}

\newcommand{\Ri}{\mathcal{R}_{i}}
\newcommand{\dRi}{\dot{\mathcal{R}}_i}
\newcommand{\Rw}{\mathcal{R}_{w}}

\newcommand{\rfree}{\mathcal{R}_{f}}

\newcommand{\pr}{p_{r}}
\newcommand{\nH}{n_{\rm H}}

\newcommand{\nHbar}{\overline{n}_{\rm H}}
\newcommand{\mH}{m_{\rm H}}
\newcommand{\muH}{\mu_{\rm H}}
\newcommand{\sfe}{\varepsilon_{*}}
\newcommand{\rhobar}{\bar{\rho}}

\newcommand{\fwind}{f_{\rm wind}}

\newcommand{\Mcl}{M_{\rm cl}}
\newcommand{\Mst}{M_*}
\newcommand{\Rcl}{R_{\rm cl}}

\newcommand{\ci}{c_i}

\newcommand{\vw}{{\cal{V}}_w}

\newcommand{\pc}{\mathrm{\, pc}}
\newcommand{\pcc}{\mathrm{\, cm}^{-3}}
\newcommand{\yr}{\mathrm{\, yr}}
\newcommand{\Myr}{\mathrm{\, Myr}}
\newcommand{\nm}{\mathrm{\, nm}}
\newcommand{\kms}{\mathrm{\, km\, s^{-1}}}
\newcommand{\Kel}{\, \mathrm{K}}
\newcommand{\Msun}{\, \mathrm{M}_\odot}

\newcommand{\vout}{v_{\rm out}}
\newcommand{\voutavg}{ \left\langle \vout \right\rangle}
\newcommand{\Abub}{A_w}

\newcommand{\clump}{\left\langle n^2 \right\rangle/\left\langle n\right\rangle^2}
\newcommand{\clumpi}{\mathfrak{C}_i}

\newcommand{\strom}{Str\"{o}mgren }

\newcommand{\WOH}{$\texttt{HW}$}
\newcommand{\ROH}{$\texttt{HR}$}
\newcommand{\WRH}{$\texttt{HWR}$}
\newcommand{\WOM}{$\texttt{MW}$}
\newcommand{\ROM}{$\texttt{MR}$}
\newcommand{\WRM}{$\texttt{MWR}$}

\received{XXX}
\revised{XXX}
\accepted{XXX}

\submitjournal{ApJ}

\shorttitle{Co-Evolution of Winds and Photoionized Gas II}
\shortauthors{Lancaster et al.}

\begin{document}

\title{The Co-Evolution of Stellar Wind-blown Bubbles and Photoionized Gas II:\\ 3D RMHD Simulations and Tests of Semi-Analytic Models}

\correspondingauthor{Lachlan Lancaster}
\email{ltl2125@columbia.edu}

\author[0000-0002-0041-4356]{Lachlan Lancaster}
\thanks{Simons Fellow}
\affiliation{Department of Astronomy, Columbia University,  550 W 120th St, New York, NY 10025, USA}
\affiliation{Center for Computational Astrophysics, Flatiron Institute, 162 5th Avenue, New York, NY 10010, USA}

\author[0000-0003-2896-3725]{Chang-Goo Kim}
\affiliation{Department of Astrophysical Sciences, Princeton University, 4 Ivy Lane, Princeton, NJ 08544, USA}

\author[0000-0001-6228-8634]{Jeong-Gyu Kim}
\affiliation{Korea Institute of Advanced Study, 85 Hoegi-ro, Dongdaemun-gu, Seoul 02455, Republic of Korea}
\affiliation{Division of Science, National Astronomical Observatory of Japan, Mitaka, Tokyo 181-0015, Japan}

\author[0000-0002-0509-9113]{Eve C. Ostriker}
\affiliation{Department of Astrophysical Sciences, Princeton University, 4 Ivy Lane, Princeton, NJ 08544, USA}
\affiliation{Institute for Advanced Study, 1 Einstein Drive, Princeton, NJ 08540, USA}

\author[0000-0003-2630-9228]{Greg L. Bryan}
\affiliation{Department of Astronomy, Columbia University,  550 W 120th St, New York, NY 10025, USA}
\affiliation{Center for Computational Astrophysics, Flatiron Institute, 162 5th Avenue, New York, NY 10010, USA}

\begin{abstract}
In a companion paper (\paperi) we presented a Co-Evolution Model (CEM) in which to consider the evolution of feedback bubbles driven by massive stars through both stellar winds and ionizing radiation, outlining when either of these effects is dominant and providing a model for how they evolve together. Here we present results from three-dimensional radiation magneto-hydrodynamical (RMHD) simulations of this scenario for parameters typical of massive star-forming clouds in the Milky Way: precisely the regime where we expect both feedback mechanisms to matter. While we find that the CEM agrees with the simulations to within 25\% for key parameters and modestly outperforms previous idealized models, disagreements remain. We show that these deviations originate mainly from the CEM's lack of (i) background inhomogeneity caused by turbulence and (ii) time-variable momentum enhancements in the wind-blown bubble (WBB). Additionally, we find that photoionized gas acts similarly to magnetic fields \citep[as in][]{Lancaster24a} by decreasing the WBB's surface area. This causes a decrease in the amount of cooling at the WBB's interface, resulting in an enhanced WBB dynamical impact.
\end{abstract}


\section{Introduction}
\label{sec:intro}

Giant molecular clouds (GMCs) are considered to be the fundamental unit of star formation in the universe, in that nearly all observed star formation occurs in these systems \citep{McKeeOstriker07,Krumholz14}. While the formation of stars in a given galaxy is governed by processes at many different scales \citep{SomervilleDave15,NaabOstriker17} including the processes leading to the formation of such GMCs \citep{Semenov17,Jeffreson24a}, the efficiency at which the gas in these clouds is turned into stars is thought to be governed primarily by the stellar feedback from massive stars \citep{Matzner02,KMBBH19,Chevance23}.

These massive stars affect their natal clouds through many different channels, including stellar wind-blown bubbles \citep{CAK75,Weaver77,HCM09,RogersPittard13,Lancaster21a}, pressure from gas heated through photoionization \citep{Spitzer78,Whitworth79,HosokawaInutsuka06,Geen15a,Geen16,JGK18}, direct radiation pressure from UV light \citep{Draine11,Raskutti16,raskutti17,JGK16,JGK18}, and indirect or reprocessed radiation pressure \citep{KrumholzMatzner09,Fall10,Skinner15,Menon22,Nebrin24}. The question of how these processes affect their natal clouds is often asked in the literature through the lens of which is dominant in any given environment \citep{KrumholzMatzner09,Fall10}, with this dominant feedback mechanism being usually considered as the sole regulator in theoretical models. In reality, all of these mechanisms are always present and often affect one another in non-linear ways \citep{Pellegrini07,Draine11}. Even if their dynamical importance varies strongly with environment, they each will have an impact on the structure of the gas in star forming clouds and thus on the observable properties of those clouds \citep{Pellegrini20a,Pellegrini20b}. Given the abundance of recent and upcoming observations of these regions \citep{SOFIA_FEEDBACK,LVM_Science,Kreckel24_Orion,SIGNALS}, which require interpretation, there is a pressing need to improve the theoretical modeling of interacting feedback processes and their observable consequences.

To that end, in \paperi\ we presented a model for the interaction of feedback from both stellar wind-blown bubbles (WBBs) and photoionized regions (PIRs) created by Lyman Continuum (LyC) radiation. In particular, we specified the regimes of parameter space in which winds or LyC radiation should be more dynamically important and developed a semi-analytic model for how the feedback bubble (FB) resulting from both of these mechanisms evolves. Here we present three-dimensional (3D), radiation magneto-hydrodynamical (RMHD) simulations of the joint feedback of both of these mechanisms in the regime where we expect both to be important. We use these simulations for the dual purposes of (i) testing the accuracy and pitfalls of the model developed in \paperi\ and (ii) understanding the effects of the presence of LyC radiation on cooling at the WBB's interface, which we have shown in previous work determines the dynamical impact of the WBB \citep{Lancaster21a,Lancaster21b,Lancaster24a}. As we will show below, the presence of each of these feedback mechanisms impacts the other in complicated ways.

In \autoref{sec:theory} we review the results of the combined evolution models of \paperi\ and the classical solutions for the evolution of WBBs and photoionized gas. We provide details on the simulation suite we have run in \autoref{sec:simulations} before analyzing them, through both comparisons to models of \paperi\ and \citet{Lancaster24a}, in \autoref{sec:results}. In \autoref{sec:discussion} we place this work in the context of past numerical works that have investigated the feedback from both WBBs and LyC radiation. In particular, using the model of \paperi\, we explain why past works, which for the most part have focused on individual stars or low density environments, had concluded that winds are generally subdominant to LyC radiation feedback. Finally, we summarize our conclusions in \autoref{sec:conclusion}.

\section{Review of Theory}
\label{sec:theory}

This section presents a summary of the key ideas relevant to the joint feedback evolution from WBBs and the PIR. This discussion is given in much greater detail in \paperi.

\subsection{Classical Theories}
\label{subsec:theory_classic}

We begin by briefly reviewing the theoretical background of feedback from LyC radiation and stellar winds. This is meant to provide a clear comparison between the semi-analytic models developed in \paperi\ and the numerical simulations given here. We characterize the former by the rate at which LyC photons are emitted, $\Qo$, and their average energy, $h\nu_i$. We characterize winds by the mass loss rate, $\mdot$, and wind velocity, $\vw$, which can be translated into the wind's mechanical luminosity, $\Lwind = \mdot\vw^2/2$, and momentum input rate, $\pdot = \mdot\vw$. We consider the impacts of the injection of energy in these forms into a uniform background medium with density $\rhobar$ and therefore number density of hydrogen atoms, $\nHbar = \rhobar/\muH\mH$, where $\mH$ is the mass of a Hydrogen nucleus and $\muH$ is the mean molecular weight per H atom (in units of $\mH$).

Stellar winds shock as they expand into the surrounding cloud, heating the wind to $10^7-10^8\Kel$ and making it vastly over-pressurized with respect to its surroundings \citep{Weaver77,Drainebook}. The resulting WBB does mechanical work on its surroundings as it expands and sweeps up the surrounding gas. The effectiveness of this work largely hinges upon the bubble's ability to retain thermal energy in its interior \citep{Weaver77,maclow88,HCM09,Rosen14,Lancaster21a}. In the `best case scenario' for wind effectiveness, all energy is retained and the classical ``energy-driven'' solution of \citet{Weaver77} applies. If we modify this solution to allow for a constant fraction, $\theta$, of the wind luminosity to be lost to cooling \citep{ElBadry19}, it is characterized by $(1-\theta)\Lwind$ with evolution of the wind bubble's radius and the momentum that it carries given by
\begin{equation}
    \label{eq:Rweaver}
    \RWeavert(t) =
    \left( \frac{125}{154\pi}\frac{(1-\theta)\Lwind t^3}{\rhobar} \right)^{1/5} 
    \, ,
\end{equation}
\begin{equation}
    \label{eq:prweaver}
    \prWeavert (t) 
    = \frac{4\pi}{5} \left( \frac{125}{154 \pi}\right)^{4/5}
    \left((1-\theta)^4\Lwind^4 \rhobar t^7 \right)^{1/5} \, .
\end{equation}
In the regime that cooling losses are small ($\theta \ll 1$), WBBs are generally dynamically dominant over PIRs \citep{paper1}. However, recent work \citep{Lancaster21a,Lancaster21b,Lancaster21c,Lancaster24a} and observations \citep{Dunne03,Townsley03,Townsley06,HCM09,Townsley11,Lopez11,Lopez14,Rosen14} suggest that $(1-\theta) \ll 1$, i.e. that cooling losses are significant.

These cooling losses likely occur either by conductive mass-loading of the interior \citep{Weaver77,maclow88} or turbulent mixing at the WBB interface \citep{Lancaster21a,Lancaster21b,Lancaster21c,Lancaster24a}. Energy losses can be efficient enough to reduce the bubble to a ``momentum-driven'' solution, characterized by $\pdot$ with dynamical evolution given by
\begin{equation}
    \label{eq:rEC}
    \RMDa (t) = \left( \frac{3}{2\pi} \frac{\alpha_p\pdot t^2}{\rhobar}\right)^{1/4} \, ,
\end{equation}
\begin{equation}
    \label{eq:pr_EC}
    \prMDa (t)  =  \alpha_p\pdot t \, ,
\end{equation}
in which the dynamical impact of WBBs is much reduced \citep{Steigman75,Lancaster21a}. We have allowed for a constant `momentum enhancement factor,' $\alpha_p$, in the above in order to parameterize energy retention beyond the purely momentum-driven limit. The physics behind $\alpha_p$ is discussed further below as well as in \paperi\, and \citet{Lancaster24a}. In dimensional form, the wind momentum input rate per unit solar mass averaged over a Kroupa stellar initial mass function (IMF) and the first $2\,\Myr$ of evolution at solar metallicity as derived from \texttt{STARBURST99} \citep{SB99} is $\pdot/\Mst = 9.5\, \kms\,\Myr^{-1}$.

The degree to which WBBs are energy- or momentum-driven in reality is still somewhat uncertain \citep{Lancaster24a}, and real WBBs likely lie somewhat in between these two extremes, as is the case for the simulations we discuss below. We include the parameters $\theta$ and $\alpha_p$ in the above relations in order to expand the realm of applicability of these models. As we will see below, our simulations are better (but not perfectly) characterized by this latter model, with constant $\alpha_p$.

LyC radiation acts to create an over-pressurized PIR outside of the WBB by ionizing and thermally heating gas in the star-forming environment. The radiation acts to quickly ionize an initial ``\strom Sphere'' with radius
\begin{equation}
    \label{eq:RSt}
\begin{split}
    \RSt &=  \left(\frac{3\Qo}{4\pi \alpha_B \nHbar^2} \right)^{1/3} \\
    &= 10.2 \pc 
    \left( \frac{\Xi}{4.1\times 10^{46}\, {\rm s}^{-1}\,\Msun^{-1}}\right)^{1/3}\\
    &\times \left( \frac{\Mst}{10^4\, \Msun}\right)^{1/3}
    \left(\frac{\nHbar}{100\pcc} \right)^{-2/3} \, ,
\end{split}
\end{equation}
which takes place roughly over a recombination time $\trec = (\alpha_B \nHbar)^{-1} \approx 10^3\, {\rm yr}$, where $\alpha_B(T=8000\,{\rm K}) \approx 3.11 \times 10^{-13}\, {\rm cm}^{3}\, {\rm s}^{-1}$ is the case B recombination rate \citep{Drainebook}. In the above dimensional version we use $\Xi \equiv \Qo/\Mst$ as derived from the same \texttt{STARBURST99} procedure mentioned above for $\pdot$, where $\Mst$ is the total stellar mass of the cluster. As the bubble expands, its pressure drops but it retains equilibrium between photoionization and recombination by photo-evaporating gas off the interior of the FB's shell. One can use a thin-shell momentum equation along with the assumption of ionization-recombination equilibrium to derive the radial evolution of the bubble as
\begin{equation}
    \label{eq:spitzer_sol}
    \RSp(t) = \RSt \left(1 + \frac{7}{4} \frac{t}{\tdio} \right)^{4/7} \, ,
\end{equation}
where
\begin{equation}
    \label{eq:tdio_def}
    \tdio \equiv  \frac{\RSp(0)}{\dot{R}_{\rm Sp}(0)}= \frac{\sqrt{3}}{2} \frac{\RSt}{\ci}
\end{equation}
is the initial dynamical expansion time of the bubble and $\ci \approx10\kms$ is the sound speed of the ionized gas \citep{Spitzer78,HosokawaInutsuka06}. The inferred momentum carried by the bubble would naively be written as
\begin{equation}
    \label{eq:pr_spitzer1}
    p_{r,{\rm Sp}}(t) = \frac{4\pi}{3} \rhobar \RSp^3 \dot{R}_{\rm Sp} = 
    \frac{8\pi}{3\sqrt{3}} \rhobar \ci \RSt^3 
    \left(1 + \frac{7}{4}\frac{t}{\tdio} \right)^{9/7} \, .
\end{equation}
However, accounting for the fact that the mass in the interior of the bubble does not contribute to the shell mass results in 
\begin{equation}
    \label{eq:pr_spitzer_adj}
    p_{r,{\rm Sp,adj}}(t) = p_{r,{\rm Sp}}(t)
    \left(1 - \left(\frac{\RSt}{\RSp}\right)^{3/2} \right) \, ,
\end{equation}
which is zero at $t=0$ as we would expect. Similar adjustments are suggested in \citet{Haid18} and \citet{Pittard22Rad}. The implications of these two formulae are compared against simulations in \autoref{app:spitzer_momentum}. We can give the relative momentum scale injected by the Spitzer solution using the pre-factor in \autoref{eq:pr_spitzer1}, which in dimensional form is given as
\begin{equation}
    \begin{split}
        \frac{8\pi}{3\sqrt{3}}\rhobar\ci\RSt^3 &= 1.8\times 10^{5} \,\Msun \kms \\
        &\times \left( \frac{\Xi}{4.1\times 10^{46}\, {\rm s}^{-1}\,\Msun^{-1}}\right) \left( \frac{\Mst}{10^4\, \Msun}\right)\\
        &\times \left(\frac{\nHbar}{100\pcc} \right)^{-1} \left( \frac{\ci}{10\kms}\right) \, .
    \end{split}
\end{equation}

\subsection{Co-Evolution Model}
\label{subsec:theory_joint}

In \paperi\ we described how WBBs are over-pressurized with respect to the surrounding PIR at early times and used this fact to infer the radius and time at which a WBB comes into pressure balance with the surrounding PIR. For the case of a momentum-driven WBB with a momentum enhancement factor, $\alpha_p$ this radius and time are given by
\begin{equation}
    \label{eq:RE_def}
    \begin{split}
        \Req &\equiv \sqrt{\frac{\alpha_p \pdot}{4\pi \rhobar \ci^2}} \\
        &=4.74\, {\rm pc} \, \left(\frac{\alpha_p\pdot}{10^5 \, M_{\odot}\, {\rm km/s/Myr}} \right)^{1/2} \\
        &\times \left(\frac{\nHbar}{100\pcc} \right)^{-1/2} 
        \left( \frac{\ci}{10\, {\rm km/s}}\right)^{-1}\, ,
    \end{split}
\end{equation}
and
\begin{equation}
    \label{eq:tE_def}
    \begin{split}
        \teq &\equiv \frac{1}{4\pi \ci^2} \sqrt{\frac{2\pi}{3} \frac{\alpha_p \pdot}{\rhobar}} = \frac{\Req}{\sqrt{6} \ci} \\
         &= 1.89 \times 10^5 \, {\rm yr}\, 
        \left(\frac{\alpha_p\pdot}{10^5 \, M_{\odot}\, {\rm km/s/Myr}} \right)^{1/2} \\
        &\times \left(\frac{\nHbar}{100 \pcc} \right)^{-1/2} 
        \left( \frac{\ci}{10\, {\rm km/s}}\right)^{-2}\, .
    \end{split}
\end{equation}
We quantify the relative importance of WBBs and the PIR using the quantity
\begin{equation}
    \label{eq:etaMD_def}
    \begin{split}
        \ReqRSt \equiv \frac{\Req}{\RSt} 
        &= 0.47 \, \left(\frac{\Qo}{4\times10^{50} \, {\rm s}^{-1}} \right)^{-1/3} \\
        & \times \left( \frac{\alpha_p \pdot}{10^5 \, M_{\odot}\, {\rm km/s/Myr}}\right)^{1/2}\\
        &\times\left( \frac{\nHbar}{100 \pcc} \right)^{1/6}
        \left( \frac{\ci}{10 \, \kms}\right)^{-1}
        \, .
    \end{split}
\end{equation}

As shown in \paperi, $\ReqRSt \lesssim 1$ for the parameters typical of star-forming regions in the Milky Way, meaning that in general photoionized gas will not be `over-run' by the WBBs (as can be the case in denser star-forming environments) but will be significantly disturbed by the WBBs. Motivated by this fact we developed a semi-analytic co-evolution model (CEM) for the evolution of FBs driven jointly by stellar winds and LyC radiation.

This CEM consists of two distinct phases. The first ``early phase'' consists of the WBB and PIR evolving independently of one another and following their classical solutions, as detailed above, and lasting up until $\teq$. This lack of interaction at early times is motivated by (i) the fact that the WBB is over-pressurized at these times and therefore does not significantly feel the presence of the PIR and (ii) in the context of a unstratified background medium, the WBB's over-dense shell does not strongly trap ionizing radiation at early times, and therefore the PIR does not significantly feel the presence of the WBB\footnote{See Appendix A of \paperi\, for a more in depth discussion of the effects of radiation trapping in stratified background density profiles.}.

At $t\gtrsim\teq$ the CEM switches to a ``co-evolution phase'' at which point the WBB and the PIR are forced to be at the same pressure. We then subsequently solve for the evolution using a momentum equation for the mass and momentum carried by the joint FB while enforcing pressure equilibrium across the WBB interface and ionization-recombination equilibrium in the PIR.

While we derive versions of this CEM for both the momentum and energy-driven case in \paperi, we will only make use of the momentum-driven case here as it is more applicable to the simulations against which we are testing the model. For the momentum-driven case, during the co-evolution phase, the outer radius $\Ri$ of the PIR evolves following Equation 28 of \paperi, while the relationship between $\Ri$ and $\Rw$ (the outer radius of the WBB) is given by Equation 32 of that work.

The momentum-driven co-evolution model (MD-CEM) in the co-evolution phase consists of a single, second order differential equation which requires initial conditions on the wind bubble radius, $\Rw$, the ionization front radius, $\Ri$, and the velocity of the ionization front, $\dRi$. These are determined by ensuring the continuity of $\Rw$ and the total momentum across the transition between the two phases. Since pressure equilibrium is instantaneously enforced at $t= \teq$, it is required that there is a discontinuity in one parameter of the model, which we choose to be $\Ri$ (or equivalently, the density of the ionized gas $\rho_i$).

Appendix C of \paperi\ derives a dimensionless version of the MD-CEM evolution equation which leads to a family of solutions parameterized by a single value $\ReqRSt \equiv \Req/\RSt$. Figure 4 of \paperi\ presents solutions for a range of $\ReqRSt$.

\begin{deluxetable}{cccc}
\tablecaption{Parameters of simulation suite.\label{tab:sim_params}}
\tablewidth{0pt}
\tablehead{
\colhead{Name} & Dynamics & \colhead{Wind On}  & \colhead{LyC Radiation On}}
\startdata
\WOH & HD  & \checkmark  & X \\
\ROH & HD  & X           & \checkmark \\
\WRH & HD  & \checkmark  & \checkmark \\
\WOM & MHD & \checkmark  & X \\
\ROM & MHD & X           & \checkmark \\
\WRM & MHD & \checkmark  & \checkmark \\
\enddata
\end{deluxetable}

\begin{figure*}
    \centering
    \includegraphics[width=\textwidth]{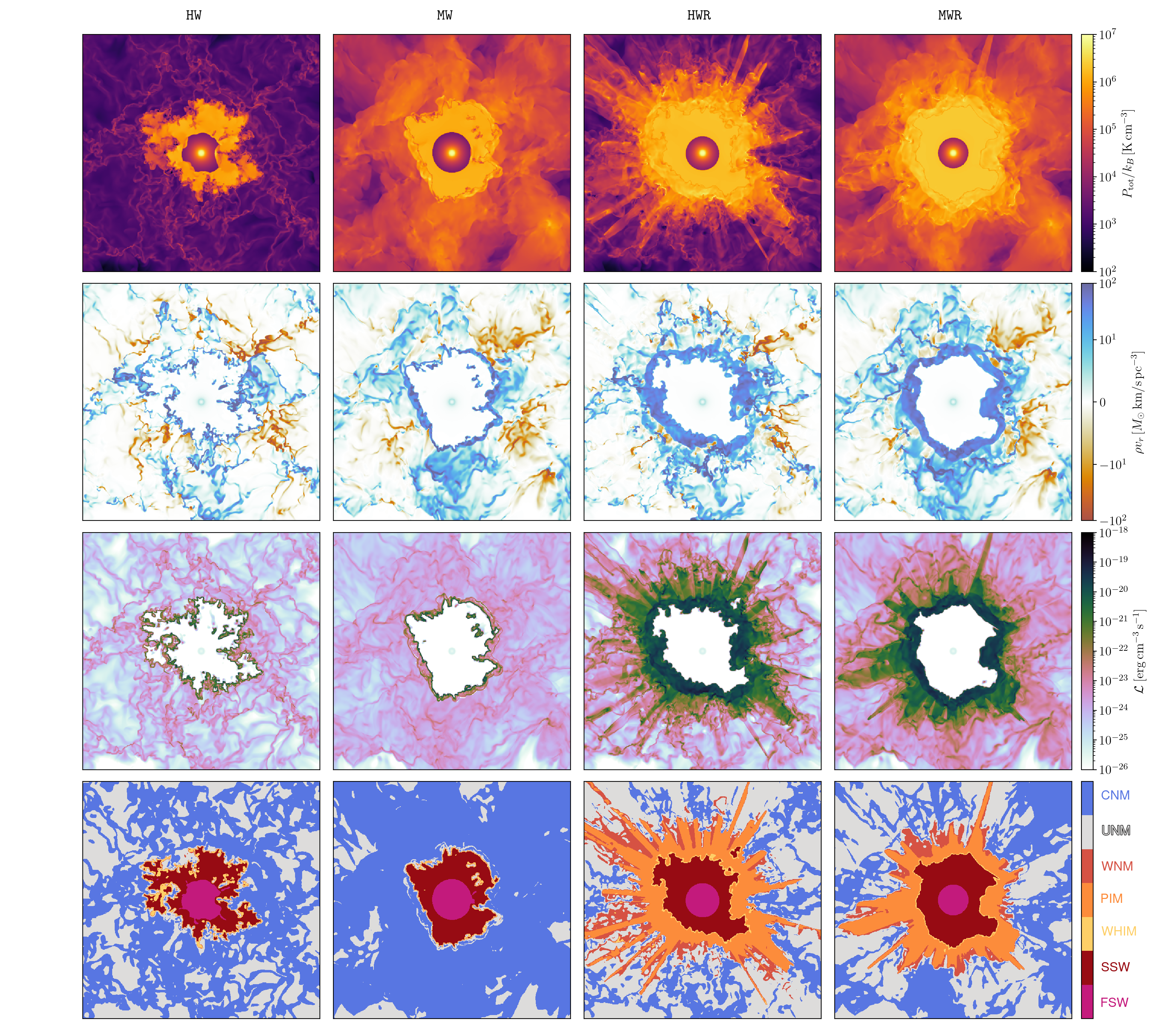}
    \caption{Slices through the $z=0$ plane of our high resolution simulations, with $\Lbox/\delx = 512$, at the time when $\Rw \approx \Lbox/6 = 6.7\pc$; this is approximately $\Req$ for $\alpha_p = 3$. The columns give snapshots from each of four simulations; from left to right these are: \WOH, \WOM, \WRH, and \WRM. The rows show slices of different physical quantities, from top to bottom these are: (i) total pressure, including thermal and magnetic terms as appropriate, (ii) radial momentum density of the gas, (iii) the total volumetric radiative cooling, and (iv) a schematic that separates the slice into different components, indicated by color. Color bars associated with these quantities are shown at the right. In the \WOM\ and \WRM\ runs the mean magnetic field points out of the page. An animated version of the figure is available \href{https://vimeo.com/1088473741/39a2cfcc61?ts=0&share=copy}{here}.}
    \label{fig:mega_plot}
\end{figure*}

\begin{deluxetable*}{ccccc}
    \tablecaption{Definitions of Gas Phases and Regions.\label{tab:phase_defs}}
    \tablewidth{0pt}
    \tablehead{
    \colhead{Phase} & \colhead{Abbreviation} & \colhead{Temperature Condition} 
    &  \colhead{Velocity Condition}& \colhead{$x_e$ Condition}}
    \startdata
    Free Stellar Wind & FSW & $T>10^6\, {\rm K}$ & $v_r > \vw /2$  &  \\
    Shocked Stellar Wind & SSW & $T>10^6\, {\rm K}$ & $v_r < \vw /2$  & \\
    Warm-Hot Ionized Medium  & WHIM & $10^4\, {\rm K}<T<10^6\, {\rm K}$ & $v_r < \vw /2$
    & \\
    Photoionized Medium & PIM & $6085\, {\rm K}<T< 10^4\, {\rm K}$ & & $x_e>0.9$ \\
    Warm Neutral Medium & WNM & $6085\, {\rm K}<T< 10^4\, {\rm K}$ & & $x_e < 0.9$ \\
    Thermally Unstable Medium & UNM & $181\, {\rm K}<T<6085\, {\rm K}$ &  & \\
    Cold Neutral \& Molecular Medium & CNM & $T<181\, {\rm K}$ &  & \\
    \enddata
    \tablecomments{Names assigned to gas components based on temperature, velocity, and $x_e$ conditions}
\end{deluxetable*}

\section{Simulations}
\label{sec:simulations}

In order to confront the various theoretical interpretations and predictions laid out in \paperi\ and reviewed in \autoref{sec:theory}, we run a suite of numerical simulations of FBs both with and without background magnetic fields and cycling the feedback effects that are included: No Feedback, Winds Only, Photoionization Only, and both Winds and Photoionization. We run our simulations with the magneto-hydrodynamics (MHD) code \textit{Athena} \citep{Stone08_Athena} supplemented with several additional physics modules that are part of the \textit{Athena}-TIGRESS code base \citep{CGK_TIGRESS1}. The radiation solver employs adaptive ray tracing \citep{AbelWandelt02}, with implementation and test laid out in \citet{JGK17}; applications to studying the effects of radiation on the evolution and destruction of star-forming clouds are presented in \citet{JGK18} and \citet{JGK21}. Our simulations are run using the linearized Roe Riemann solver \citep{Roe81} with second order, piecewise linear spatial reconstruction, and the unsplit van Leer-type integrator of \citet{StoneGardiner09}. As in \citet{Lancaster24a}, for our MHD simulations we partially employ a diffusive, first order method for constrained transport that is a version of the \texttt{UCT-HLL} method of \citet{LondrilloDelZanna04} as described in \citet{MignoneDelZanna21}. This method is only applied in the wind feedback region and in regions with strong magnetic field gradients ($\delta B/B > 10$ from cell to cell) and the diffusivity only applies to the magnetic field. Our simulations employ outflow boundary conditions. We use a CFL criterion for our MHD time step of $\Delta t_{\rm MHD} = \mathcal{C}\Delta x/ v_{\rm max}$ with $\mathcal{C} = 0.3$, $\Delta x$ the resolution, and $v_{\rm max}$ the maximum (absolute value) signal speed given by solving the Riemann problem at cell interfaces over the whole grid. We compute ray-tracing based on the above CFL time-step criterion restricted to gas with $T<2\times 10^4\,{\rm K}$. This significantly improves performance by preventing the need to re-evaluate the radiation field on the CFL time of the hot gas, which is completely unaffected by the radiation in our treatment. In the following subsections we describe various aspects of the code.

\subsection{Cloud Setup}
\label{subsec:cloud_details}

In all of the simulations presented here we choose conditions similar to that of massive GMCs in the Milky Way. Specifically, we are motivated by conditions as may be found in a cloud of total mass $\Mcl = 10^5\, \Msun$ and radius $\Rcl = 20\, \pc$, similar to the Orion A GMC. We are interested in the most massive clouds as these are likely where the majority of stars form, due to the distribution of cloud masses \citep{WilliamsMcKee97,Rice16,Tosaki17}.

Our cloud property choices imply a mean density in the simulation domain at $t=0$ of $\rhobar = 2.02 \times 10^{-22} \, {\rm g} \pcc$ or a mean number density of hydrogen nuclei $\nHbar= \rhobar/\muH\mH = 86.25\pcc$, where we have assumed the mean molecular weight per hydrogen nucleus, $\muH = 1.4$. The simulations are run in a box with side-length $\Lbox = 2\Rcl$ at varying resolution. Initially, the density is uniform throughout the simulation domain. In our simulations with magnetic fields we additionally include a uniform field $\mathbf{B} = B_0 \mathbf{\hat{z}}$ throughout the domain with $B_0 = 13.5\, \mu{\rm G}$. This choice of field strength is motivated by a dimension-less mass-to-flux ratio, $\mu_{\Phi}$, of $2$ as in the `$\alpha$-series' models of \citet{JGK21}.

In order to create a realistic background density field, we initially `stir' our background density fields as in \citet{Lancaster21b}. Specifically, we initialize a Gaussian random field with a power spectrum $|v_k| \propto k^{-2}$ with wave-number $k$ for $2 \leq k \Lbox/2\pi \leq 64$. We normalize the magnitude of the velocity field so that the kinetic energy per unit mass in the domain, $\tilde{E}_K$ is twice what the gravitational potential energy per unit mass, $\tilde{W}$, would be in a GMC of $\Mcl = 10^5 \, \Msun$ and $\Rcl = 20\, {\rm pc}$. We then allow the turbulence to decay until $\tilde{E}_K=\tilde{W}$. In our MHD runs, this initial turbulent evolution acts to effectively tangle our initially uniform magnetic field. The isotropy of the 3D bubble morphologies of our MHD runs (see Figure 5 of \citet{Lancaster24a}) indicate that the background field does not maintain a directional preference for the $\hat{\mathbf{z}}$ direction it is initialized in.

We have chosen our initial conditions to be representative of the large scale properties of a GMC: the background into which the FB expands is inhomogeneous and turbulent, but statistically homogeneous and isotropic. That is, there is no preferred direction on large scales. We make this choice as we are particularly concerned with the evolution of the FB driven by the accumulation of feedback energy from many massive stars in the cloud. While this provides a fairer point of comparison to the CEM developed in \paperi, it does not accurately represent the FBs from individual massive stars. In \paperi\, we argue that these individual star FBs should relatively quickly percolate into a cluster-driven FB.

\subsection{Cooling Physics}
\label{subsec:cooling}

We include non-equilibrium (photo-)chemistry and heating/cooling for solar metallicity conditions mostly as described in \citet{JGK_NCR23}. The cooling and chemistry are computed in an operator split manner using sub-cycling of the relevant chemical species and cooling equations. The sub-cycle time-step is chosen as $\Delta t_{\rm sub} = {\rm min}\left(0.1 \times \min_j t_j,\, \Delta t_{\rm MHD}\right)$. That is, the sub-cycle is no longer than one tenth of the minimum reaction time, where the $j$ index runs over all reactions (species abundance changes and cooling/heating). We employ a slightly earlier version of the code base than that described in \citet{JGK_NCR23}, with important differences noted below.

We follow the non-equilibrium abundance of hydrogen species (H, H$^+$, and H$_2$) and equilibrium abundance of carbon- and oxygen-bearing species (C, C$^+$, O, O$^+$, CO). The carbon and oxygen species we follow are key coolants of neutral (atomic and molecular) gas, with equilibrium abundances computed based on the local UV radiation field following \citet{Gong17}. The heating and cooling rates for cold and warm gas ($T < 2\times 10^4\,{\rm K}$) are calculated using local density, temperature, species abundances, radiation energy density, and cosmic ray ionization rate. For cooling (by helium and metals) in hot gas ($T > 3.5\times 10^4\,{\rm K}$), we adopt the ion-by-ion collisional-ionization equilibrium (CIE) cooling rates of \citet{GnatFerland12} tabulated as a function of temperature. For cooling at intermediate temperature ($2.0\times 10^4\,{\rm K}<T<3.5\times 10^4\,{\rm K}$), we make a smooth transition between non-equilibrium cooling and the CIE cooling using a sigmoid function.  The hydrogen cooling (e.g., Ly$\alpha$, free-free, radiative-recombination, H$_2$ ro-vibrational) is calculated across the full temperature range using non-equilibrium abundances.

For cooling in the PIM, both this work and the work of \citet{JGK_NCR23} explicitly track the cooling contribution from C$^+$, though this is generally only an important coolant in neutral gas. Our approach to other nebular line cooling differs from that of \citet{JGK_NCR23}. \citet{JGK_NCR23} approximate other nebular line cooling by fitting a functional form to the cooling function calculated using CMacIonize \citep{VW18} for fixed ionization states of O, N, Ne, and S. In our simulations we approximate this nebular line cooling by calculating the O$^+$ abundance as $x_{{\rm O}^+} = x_{{\rm H}^+}$ as the ionization potentials are very similar \citep{Drainebook}. We then calculate the cooling due to collisional excitation of the [OII] fine-structure lines and multiply this by a factor of 4 to approximate the total nebular cooling (other than from C$^+$). We have validated in post-processing that this gives a very similar level of cooling to the nebular line cooling treatment of \citet{JGK_NCR23}.

For physical conditions considered in our simulations, the main coolants are C$^+$, O, grain-assisted recombination, and Ly$\alpha$ in cold and warm neutral gas, the approximate nebular line emission in warm photoionized gas, and thermal bremsstrahlung emission and line emission from metal ions in shocked hot winds.

At the WBB surfaces of our wind-only simulations (also presented in \citet{Lancaster24a}), collisionally ionized gas shares an interface directly with neutral gas, leading to numerical diffusion of $x_{\rm H}$ (the neutral Hydrogen fraction) into hot ($T=10^4-10^{5.5}\, {\rm K}$) gas. This leads to an excess of cooling due to Ly$\alpha$ that otherwise would not be present in this gas. Since other coolants from ionized species in this temperature range are also very strong, and since we already demonstrated in \citet{Lancaster24a} that the cooling in these interfaces is not properly resolved, this does not detract from any of the main results presented in this work or \citet{Lancaster24a}. Indeed, cooling associated with numerical diffusion of neutral gas is not important in the simulations including both winds and LyC radiation (where collisionally ionized gas interfaces with photoionized gas) which are the key points of comparison for this work.

A similar effect also takes place at the interface between warm ($T=10^3-10^4\,{\rm K}$) gas and cold, molecular gas. There, numerical diffusion of $x_{{\rm H}_2}$ into this warm gas leads to an excess of ro-vibrational H$_2$ cooling. While this cooling dominates at these temperatures when it should not, the cooling at these temperatures is overall unimportant for the dynamics of the FBs in either scenario (with or without LyC radiation). Finally, neither of these effects should be important in simulations presented using the \citet{JGK_NCR23} work \citep{TIGRESS_NCR,TIGRESS_NCRZ,Linzer24} as these works employed cuts on $x_{\rm H}$ and $x_{{\rm H}_2}$ for Ly$\alpha$ and ro-vibrational cooling which largely mitigated these effects.

For the background heating, we include FUV heating via the grain photoelectric effect. We assume the uniform mean intensity at FUV wavelengths with the density-dependent attenuation applied as $J_k = J_k^{\rm Draine} e^{-\tau_k}$. Here $J_k^{\rm Draine}$ is \citet{Draine78}'s interstellar radiation field $J_{\rm LW}^{\rm Draine} = 0.3\times 10^{-4}\,{\rm erg}\,{\rm s}^{-1}\,{\rm cm}^{-2}\,{\rm sr}^{-1}$ and $J_{\rm PE}^{\rm Draine} = 1.8\times 10^{-4}\,{\rm erg}\,{\rm s}^{-1}\,{\rm cm}^{-2}\,{\rm sr}^{-1}$ for Lyman-Werner (LW) and photoelectric (PE) bands, respectively (defined in the second paragraph of \autoref{subsec:feedback_description} just below); $\tau_k = \nH \sigma_{{\rm d},k} L_{\rm shld}$ with dust absorption cross section per H $\sigma_{\rm d,k}$ and density-dependent local shielding length $L_{\rm shld} = 5\,{\rm pc}(\nH/10^2\pcc)^{-0.7}$ \citep{JGK21}. We also adopt a constant primary CR ionization rate of $2.0\times 10^{-16}\,{\rm s}^{-1}$, which is the dominant source of ionization in the neutral gas.

\subsection{Feedback Physics}
\label{subsec:feedback_description}

We use the `hybrid' wind injection method as described in \citet{Lancaster21b}. In this method, mass and energy are injected within an `injection region.' The relative amount of thermal/kinetic energy that is injected is smoothly interpolated between purely thermal injection at the center of the injection region and purely kinetic energy at the edge of the injection region. The injected wind mass is tracked via a passive scalar field $\rho_{\rm wind}$ which we use to define the mass fraction in wind material $\fwind \equiv \rho_{\rm wind}/\rho$. We also apply the first-order flux correction in the wind injection region.

For our runs with radiative feedback, we use the adaptive-ray tracing (ART) implementation of \citet{JGK17}, with ray-splitting ensuring that at least four rays from the source intersect each resolution element. We perform radiative transfer in three separate bands: (i) the photo-electric (PE; $110.8\nm < \lambda \le 206.6\nm$) band, (ii) the Lyman-Werner (LW; $91.2\nm < \lambda \le 110.8\nm$) band and, (iii) the Lyman Continuum (LyC; $\lambda \le 91.2\nm$) or Hydrogen-ionizing band. The PE and LW bands are primarily responsible for heating of the background through the photo-electric effect on dust grains and the photo-dissociation of ${\rm H}_2$ and ${\rm CO}$, respectively. The Lyman-Continuum band is the main band of dynamical interest for this work. We assume an average photon energy of $8.4,\, 12.2,\, 20\, {\rm eV}$ in the PE, LW, and LyC bands respectively. These energies are averages over the given band and over the first 2 Myr of evolution for a simple stellar population with Kroupa IMF and solar metallicity as calculated with the \texttt{Starburst99} (SB99) code \citep{SB99,Leitherer14} (see also Figure 22 and Table 3 of Appendix B in \citet{JGK_NCR23}).

In this work we ignore absorption of LyC photons by dust in the PIR by setting $\sigma_{\rm d, LyC} = 0$. This is consistent with the assumptions made in \paperi\footnote{While dust composition and absorption in HII regions remains incompletely understood, harsh UV radiation fields tend to lower grain abundance by destroying PAHs and expelling grains \citep{Akimkim17,Egorov23}. We defer self-consistent modeling of the effect of PAH destruction and gas-dust coupling to future work.}. Additionally, we do not treat the scattering of UV light by dust grains. Since dust is thought to have low albedo ($\sim 0.2-0.4$) and is generally forward scattering ($\left\langle \cos \theta_{\rm sca}\right\rangle\gtrsim 0.6$) in the UV this should not have a large impact on our results \citep{JGK_NCR23,WeingartnerDraine01,HensleyDraine23}. Given that we do not intend to simulate the effects of dust, which provide the main source of diffusion to UV photons, the ART method we employ (and the high angular resolution it affords) is especially well-suited to modeling the dynamical impact of LyC radiation which interacts with neutral gas, which has a very high absorption cross-section.

After creating the density inhomogeneity in the background through turbulence, as described in \autoref{subsec:cloud_details}, we place a source of constant luminosity wind and/or radiative feedback at the center of the domain. Throughout the paper, the onset of feedback will be referred to as $t=0$. We choose the constant values of $\Lwind$, $\mdot$, and $\Qo$ used in our simulations by averaging these quantities over the first $2\,\Myr$ of evolution as dictated by SB99 \citep{SB99,Leitherer14} for the `standard' output at solar-metallicity using a Kroupa IMF \citep{KroupaIMF} along with the Geneva, non-rotating stellar tracks \citep{Ekstrom12}.  This gives us a wind mechanical luminosity per unit stellar mass of $\Lwind/M_* = 9.75 \times 10^{33} \,{\rm erg}\,{\rm s}^{-1}\, \Msun^{-1}$, a mass loss rate per unit stellar mass of $\mdot/M_* = 2.965\times 10^{-3} \, \Myr^{-1}$, a luminosity per unit solar mass of $\mathcal{L}_{\rm PE}/M_* = 318.42 \, L_{\odot}\Msun^{-1}$, and $\mathcal{L}_{\rm LW}/M_* = 148.57 \, L_{\odot}\Msun^{-1}$ in the PE and LW bands respectively, and an LyC photon emission rate per unit stellar mass of $\Xi \equiv \Qo/M_* = 4.106\times 10^{46} \,{\rm s}^{-1}\, \Msun^{-1}$. The source particle has an equivalent mass of $M_* = 5\times 10^3 \, \Msun$, which determines the wind and radiation luminosities. We do not include the effects of direct radiation pressure feedback (radiation provides no source term to the momentum equation), in order to create a better comparison against the theory developed in \autoref{sec:theory}, which does not include the effects of radiation pressure. Radiation pressure is subdominant to effects from photoionized gas in the regime under consideration here \citep{JGK18, Fukushima20}.

\subsection{Simulation Suite}
\label{subsec:suite_description}

We run 6 different types of simulation corresponding to different choices along two axes:
\begin{itemize}
    \item[(1)] The inclusion or exclusion of magnetic fields (\texttt{H} or \texttt{M} respectively).
    \item[(2)] The choice of feedback: stellar wind feedback alone, radiative feedback alone, or both (\texttt{W}, \texttt{R}, or \texttt{WR} respectively).
\end{itemize}
\autoref{tab:sim_params} provides a list of these choices. The simulations without LyC radiation (\WOH\ and \WOM) are identical to those presented in \citet{Lancaster24a}. The \WOH\ and \WOM\ simulations are discussed here only insofar as they are used to assess the difference in dynamical impact of WBBs with and without LyC radiation in \autoref{subsec:momentum} and \autoref{subsec:PIR_cooling_effect}. The radiative feedback only simulations (\texttt{HR} and \texttt{MR}) are discussed only in \autoref{app:spitzer_momentum}. The simulations that contain both WBBs and LyC radiative feedback (\WRM\ and \WRH) make up the main part of the analysis of this work.

\begin{figure*}
    \centering
    \includegraphics[width=\textwidth]{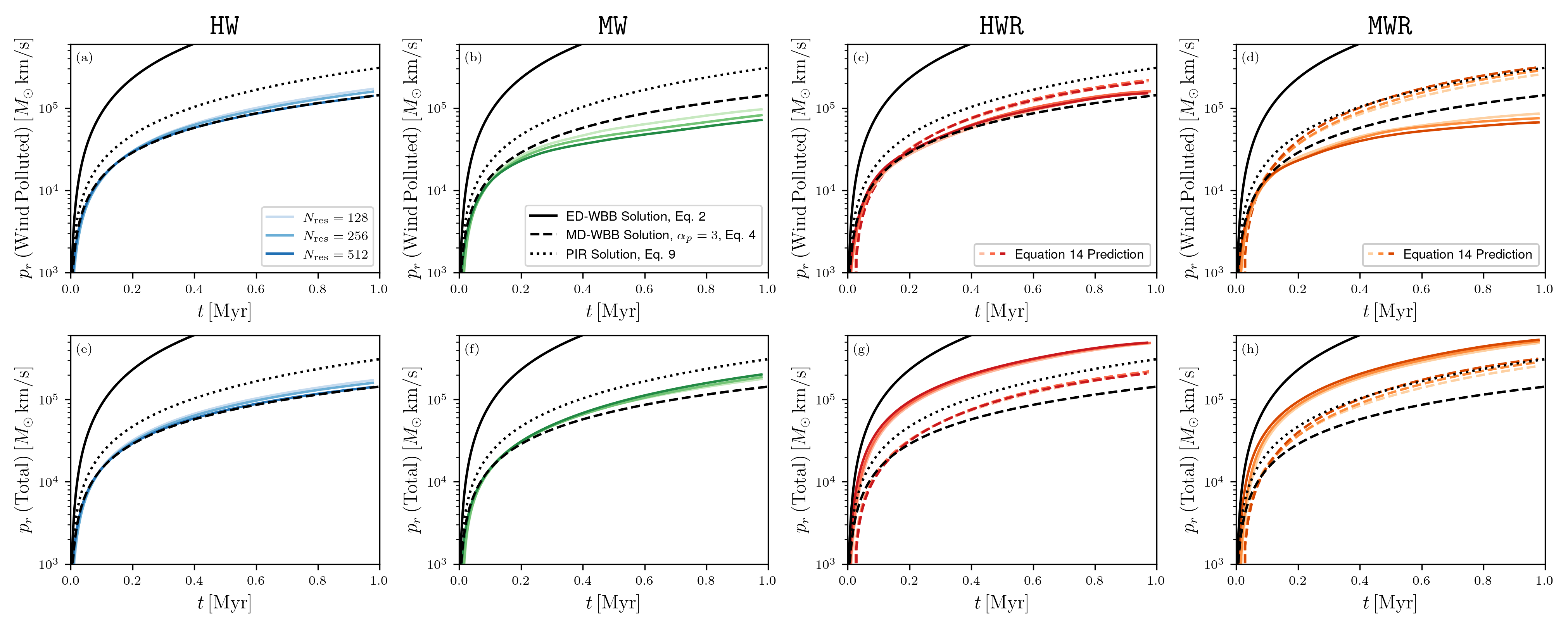}
    \caption{Radial momentum measured in each of our simulations compared to various analytic and semi-analytic predictions. Each panel shows the momentum for the classical energy-driven (\autoref{eq:prweaver}, black solid) and momentum-driven (\autoref{eq:pr_EC}, with $\alpha_p = 3$, black dashed) along with the classical PIR (\autoref{eq:pr_spitzer_adj}, black dotted) evolution. Columns from left to right represent results from the \WOH, \WOM, \WRH, and \WRM\, simulations respectively. The solid, colored lines in each panel show momentum as measured in the simulations using two different methods: top row panels show momentum measured in `wind-polluted', $\fwind > 10^{-4}$, gas, bottom row panels show total momentum as measured against reference simulations without feedback. The dashed, colored lines in panels (c), (d), (g), and (h) represent the total momentum input by the WBB, measuring its dynamical impact using \autoref{eq:alphap_derive} (appropriately time-integrated, see text). These lines in panels (c) and (g) are identical, as are those in panels (d) and (h). In each panel the variation in color of the lines indicates varying resolution, darker colored lines indicate higher resolution.}
    \label{fig:momentum_comp}
\end{figure*}

Each of these simulations are run at three different spatial resolutions, $\Nres \equiv \Lbox/\delx = 128,\, 256,\,$ and $512$, for approximately $1\Myr$ in time after the onset of feedback. A slice through the $z=0$ plane for our highest resolution simulations with stellar wind feedback is shown in \autoref{fig:mega_plot} for the time at which $\Rw \approx \Rcl/3 = \Lbox/6$. In particular we display slices in the (i) total pressure, including thermal and magnetic pressure terms as appropriate, (ii) the radial momentum density of the gas, measured with respect to the feedback source, (iii) the total volumetric rate of radiative cooling, and (iv) a breakdown of the domain into distinct phases, which are listed in \autoref{tab:phase_defs}. Here, we intend WHIM to mean `warm-hot ionized medium' analogous to the `Ionized' phase of \citet{Lancaster21b}. Other differences in phase definitions between \citet{Lancaster21b} and this work are mostly due to the change in the module we are using for cooling and heating physics and are compatible with the choices made in \citet{Lancaster24a} with additional phases included for the analysis of the PIR. To that end, we separate between the warm neutral medium and the photoionized medium using the abundance of free electrons, $x_e \equiv n_e/\nH$, where $n_e$ is the number density of free electrons.

For each choice of resolution and choice of hydrodynamics solver (pure hydrodynamics or MHD) we also run separate simulations which include no feedback but just evolve the background turbulence. We refer to these as the `no feedback' simulations and we use them for comparison purposes in the next section.

\section{Analysis and Results}
\label{sec:results}

In analyzing the results of our simulations in this section we have two primary goals (i) to assess the impact of the presence of the PIR on the picture of WBB dynamics briefly described in \paperi\ and further laid out recently in \citet{Lancaster24a} and (ii) to test the various predictions of the CEM laid out in \paperi\ and briefly reviewed in \autoref{sec:theory}. To this end we will begin by describing the total radial momentum evolution in \autoref{subsec:momentum} in both the wind-only and wind + LyC radiation simulations.
In presenting predictions associated with the ionized gas we take an ionized gas sound speed of $\ci = 10\, {\rm km/s}$, consistent with an ionized gas temperature of $T_i = 7269\Kel$.

\subsection{Momentum Evolution}
\label{subsec:momentum}


\begin{figure*}
    \centering
    \includegraphics[width=\textwidth]{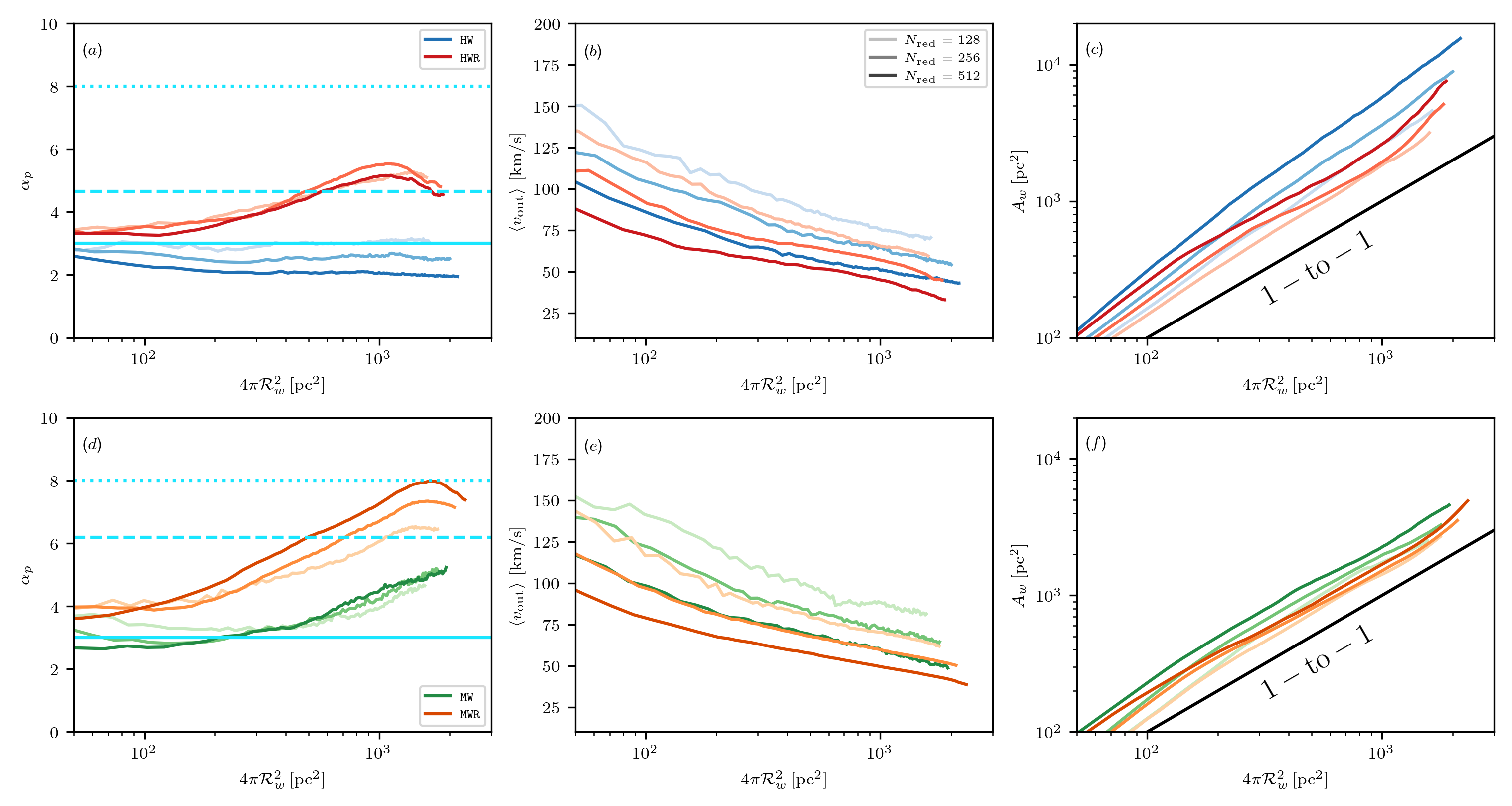}
    \caption{A comparison of the factors affecting the dynamical impact of WBBs (similar to Figure 10 of \citet{Lancaster24a}) between simulations with and without LyC radiation. The x-axis of each panel indicates the sphere-equivalent area of the WBB. \textit{Top Panels (a-c)}: Comparison of \WOH\, (blue) and \WRH\, (red) simulations. \textit{Bottom Panels (d-f)}: Comparison of \WOM\, (green) and \WRM\, (orange) simulations. In all panels, darker lines indicate higher resolution. \textit{Left panels (a,d)}: The momentum enhancement factor, $\alpha_p$, of the WBB as calculated using \autoref{eq:alphap_derive}. Lines of constant $\alpha_p=3,\, 8$ are shown as bright blue solid and dotted lines respectively, lines of $\alpha_p= 4.66$ (panel (a)) and $\alpha_p = 6.20$ (panel (d)) are shown as bright blue dashed lines. These values correspond to averages over time in the simulations with LyC radiation. \textit{Middle Panels (b,e)}: The effective dissipation velocity at which high-enthalpy gas is moved into the WBB interface. \textit{Right Panels (c,f)}: The surface area of the WBBs. The 1-to-1 lines of $\Abub = 4\pi \Rw^2$ are shown in black.}
    \label{fig:wind_evol_comp}
\end{figure*}

In \citet{Lancaster21b} we measured the total momentum carried by the WBB by taking the total momentum carried in `wind-polluted gas,' defined as any grid-cell with a fraction of its mass coming from wind material, $\fwind>10^{-4}$. In wind-only simulations this is an accurate measure of momentum imparted by the bubble since turbulent mixing and numerical diffusion across the wind bubble's interface allow for gas in the thin-shell of the WBB (which carries the vast majority of the momentum) to be precisely selected.

When magnetic fields are included, as in \citet{Lancaster24a}, momentum is able to be carried forward, out of `wind-polluted' gas, by fast magneto-sonic waves. This results in a thick, magnetized shell which carries the majority of the imparted momentum \citep[see e.g.][]{OffnerLiu18}, as is visible in the momentum density slice of the \WOM\, run in \autoref{fig:mega_plot}. In order to better track the total momentum imparted by the bubble in these simulations, we measure the total radial momentum in the domain and compare it against an identical simulation, with the same turbulent initial conditions, but with no feedback. This allows us to track \textit{all} momentum imparted by the FB, including by the PIR in simulations that include LyC radiation.

As a first investigation of the momentum impact of the various FBs, we present both measurements of total momentum in \autoref{fig:momentum_comp}. In the top panels we show momentum measured in wind polluted gas, with $\fwind > 10^{-4}$ (as in \citet{Lancaster21b}), while in the bottom panels we show total momentum compared to reference simulations without any feedback (as in \citet{Lancaster24a}) both as solid colored lines. We can see from the agreement between the top and bottom panels in the \WOH\, simulations that these measures are virtually identical in this case. In fact, they agree within 10\% throughout the simulation for every resolution, and are in better agreement at higher resolution. On the other hand, there is significant deviation in the \WOM\, simulations, where we see a decrease in wind-polluted momentum (top panel) at higher resolution, and an opposite trend in the total momentum from comparison (bottom panel). This is indicative of the proposed mechanism that MHD waves carry momentum out of wind-polluted gas: enhanced numerical diffusion at lower resolution allows the `wind-polluted' gas to cover more of the thick, magnetized shell which carries all of the momentum.

Regarding the simulations with photoionization, in the panels at the right of \autoref{fig:momentum_comp}, we see that the $\fwind$ based measurements indicate broad agreement with the $\alpha_p =3$, MD theory for the \WRH\, simulations, while the \WRM\, simulations indicate a similar dip in momentum as seen in the \WOM\, simulations. In both simulations, momentum measurements made using the comparison to a reference simulation show momentum injection far above the MD wind theory, indicative of the dynamical impact of the PIR pressure, represented by the black dotted line (\autoref{eq:pr_spitzer_adj}). While we would like to confront these measurements with the theory developed in \paperi, we would first like to isolate the impact that the presence of the PIR has on the dynamics of the WBB. This is even a concern for the \WRH\, simulations where, while we may see good agreement between the theory and the wind-polluted measurement, there may still be a similar diffusion of momentum out of wind-polluted gas at late times when the wind-bubble's expansion becomes subsonic in the PIR. In particular, the solid lines in panel (c) of \autoref{fig:momentum_comp} may still be missing some momentum that is imparted by the wind.

In order to assess the momentum imparted by the wind in the simulations with PIR, we measure $\alpha_p$ using the theory of \citet{Lancaster24a}, in particular we use \footnote{Though we do not show it here, we have verified that the correspondence between properties of the bubble interface and the pressure of the hot gas (Equation 11 of \citet{Lancaster24a}), an integral part of the application of \autoref{eq:alphap_derive}, still hold in our simulations with LyC radiation.}
\begin{equation}
    \label{eq:alphap_derive}
    \alpha_p = \frac{3}{4} \frac{\vw/4}{\voutavg} \frac{4\pi \Rw^2}{\Abub} \, ,
\end{equation}
where $\Abub$ is the area of the interface of the WBB and
\begin{equation}
    \label{eq:vout_def}
    \voutavg \equiv \Abub^{-1} \int_{\rm \Abub} \left(\mathbf{v} - \mathbf{W}\right) \cdot \nhat\,  dA \, 
\end{equation}
is the average velocity at which gas is moved out of the bubble and into the interface.

We measure the quantities $\Abub$ and $\voutavg$ exactly as in \citet{Lancaster24a}, as specified in Section 3.3 of that work. To briefly review, $\Abub$ is measured using the \texttt{marching\_cubes} algorithm available in \texttt{scikit-image} by selecting iso-temperature surfaces at $T= 10^6\, {\rm K}$ and measuring them at the grid-scale. $\voutavg$ is measured by interpolating the gas velocity field to this surface, taking a dot product with the normal to the surface, and averaging over the surface. To account for the relative velocity of the wind interface, as given by the $\mathbf{W}$ term in \autoref{eq:vout_def}, we subtract out the time derivative of the bubble's effective radius, $d\Rw/dt(4\pi \Rw^2/\Abub)$. The inclusion of this last term, while formally more correct, has no material impact on the inference of $\voutavg$ or $\alpha_p$ as $\vout\gg W$ at the surface in all instances. \autoref{eq:alphap_derive} then gives us $\alpha_p (t)$. We use this to derive the momentum that is realistically imparted by the wind according to these measurements as $p_r = \int \alpha_p(t) \pdot dt$.

The results of this calculation are shown in panels (c), (d), (g), and (h) of \autoref{fig:momentum_comp} as the dashed colored lines. We see that in the case of the \WRH\, simulations these estimates are quite close (within 30\%) to those based on measurements from wind-polluted gas, indicating that the effect of momentum redistribution by sound waves in the PIR is less efficient than movement by magneto-sonic waves, as in the \WOM\, simulations. Differences are more pronounced in the \WRM\, simulations, presumably for the same reason as in the \WOM\, runs. In panels (g) and (h) we see by comparing the colored dashed lines (estimates of total WBB injected momentum) to the black dotted lines (analytic PIR solution of \autoref{eq:pr_spitzer_adj}) that the contributions of each mechanism to the total momentum (solid colored lines) are equal to within roughly 30\% for the \WRH\, case (PIR more important) and roughly equal in the \WRM\, models.

\subsection{Effect of the PIR on WBB Dynamics}
\label{subsec:PIR_cooling_effect}

In the theory of \citet{Lancaster24a} the dynamics of WBBs are affected by properties of cooling at their interfaces in two distinct ways: dissipation and geometry. The former controls how quickly energy can be lost at a given portion of the WBB interface whereas the latter dictates how much interface is available to dissipate and cool over. While it was shown in \citet{Lancaster24a} that the presence of magnetic fields drastically changed the geometry of the WBB interfaces, the properties of (numerically driven) dissipation at each part of the interface remained largely similar. 

In \autoref{fig:wind_evol_comp} we display a comparison of these effects in our simulations with LyC radiation as a function of the sphere-equivalent area $4\pi\Rw^2$. In particular we show $\voutavg$ and $\Abub$, measured as described in \autoref{subsec:momentum} in the middle and right panels respectively. In the left panels we use these measurements along with \autoref{eq:alphap_derive} to derive the momentum enhancement factors for these simulations and also show reference values in bright blue for $\alpha_p = 3,\, 8$ (solid and dotted lines) and average values over the simulations with LyC radiation of $\alpha_p=4.66,\, 6.20$ for the \WRH\ and \WRM\ simulations respectively in dashed lines (average values of $\alpha_p$ across time and resolution for are $\alpha_p = 2.55,\, 4.09$ for the \WOH\ and \WOM\ simulations respectively). We will use these values for our model comparisons in \autoref{subsec:joint_sim_comp}.

As we mentioned in \autoref{subsec:cooling}, the presence of the PIR at the interfaces of the WBBs in our simulations with LyC radiation means that winds no longer interface directly with neutral gas and that therefore their ability to cool through collisional excitation of Ly$\alpha$ is reduced. This results in somewhat less efficient cooling at the WBB interfaces which, in the language of \citet{Lancaster24a}, means less dissipation of energy at the interface as quantified by the average velocity at which gas is moved out of the bubble and into the interface\footnote{The translation between decreased cooling and decreased dissipation is a complex numerical process discussed in Section 4.5 of \citet{Lancaster24a}.}, $\voutavg$. This is apparent in panels (b) and (e) of \autoref{fig:wind_evol_comp}, where we see consistently smaller values of $\voutavg$ in simulations with LyC radiation compared to those without.

It is also apparent from panels (c) and (f) of \autoref{fig:wind_evol_comp} that the surface area of the WBBs in simulations with LyC radiation is lower than those without at fixed size (as quantified by the comparison to $4\pi \Rw^2$). In \citet{Lancaster24a} we explained how a similar decrease in surface area between the \WOH\, and \WOM\, simulations was due to the presence of a strong magnetic field in the swept-up shell of the \WOM\, WBB halting dynamical instabilities that create smaller scale structure and enhance $\Abub$. A similar effect is at play with LyC radiation: the resulting increase in the sound speed of the medium creates a thicker shell which halts the effectiveness of so-called thin-shell instabilities \citep{Vishniac83}. Heating due to LyC radiation also impacts the gas density structure in the background medium. In particular, the near immediate heating of all gas surrounding the WBB to an almost uniform temperature means that over-densities are also over-pressurized. These over-densities are then smoothed out on approximately a sound crossing time of the size of the over-density, which can be quite short. This effect means that the medium into which the WBB expands is much more uniform and therefore less susceptible to surface area enhancements due purely to the multi-scale density structure of the background. In \autoref{app:supp_analysis} we provide supplementary analysis of the clumping of the background and photoionized gas (\autoref{subapp:clumping}) and the thickness of the shell surrounding the WBB (\autoref{subapp:shell_thick}) that supports the claims laid out above.

The exact values of $\Abub$ and $\voutavg$ (and therefore $\alpha_p$) in reality will likely be different from those in our simulation. This is due both to differences in any surrounding geometry (particular to each cloud) and in the detailed diffusion and cooling physics at the interface, which are difficult to resolve in these simulations \citep{Lancaster24a}. This can be addressed in future work by running suites of simulations with differing geometry and developing sub-grid models to ensure that the diffusion physics at these interfaces are properly accounted for.

However, the trends in $\voutavg$ and $\Abub$ due to the inclusion of LyC radiation are both physically well motivated. It is therefore quite reasonable to conclude that the presence of LyC radiation acts to enhance the dynamical impact of WBBs, as demonstrated by the larger $\alpha_p$ values in panels (a) and (d) of \autoref{fig:wind_evol_comp}.

\subsection{Comparison to Co-Evolution Model}
\label{subsec:joint_sim_comp}

We now focus on our simulations containing LyC radiation and winds and use these as test beds for the theory developed in \paperi. The theory of \paperi\ assumes a constant $\alpha_p$ which we can see is not necessarily the case in the simulations (see \autoref{fig:wind_evol_comp}). In subsequent figures we therefore compare to models that assume $\alpha_p = 3$ and $8$ to bracket the space of realized momentum enhancement factors as well as average values over both time and resolution for the \WRH\, and \WRM\, simulations, giving $\alpha_p = 4.66,\, 6.20$ respectively. For quantitative comparisons given in the text we compare the models at a given resolution to the CEM with an $\alpha_p$ value chosen to match the average value realized in that simulation, not additionally averaged across resolutions. These average values are provided in the second column of \autoref{tab:cem_comp}.

\begin{figure*}
    \centering
    \includegraphics{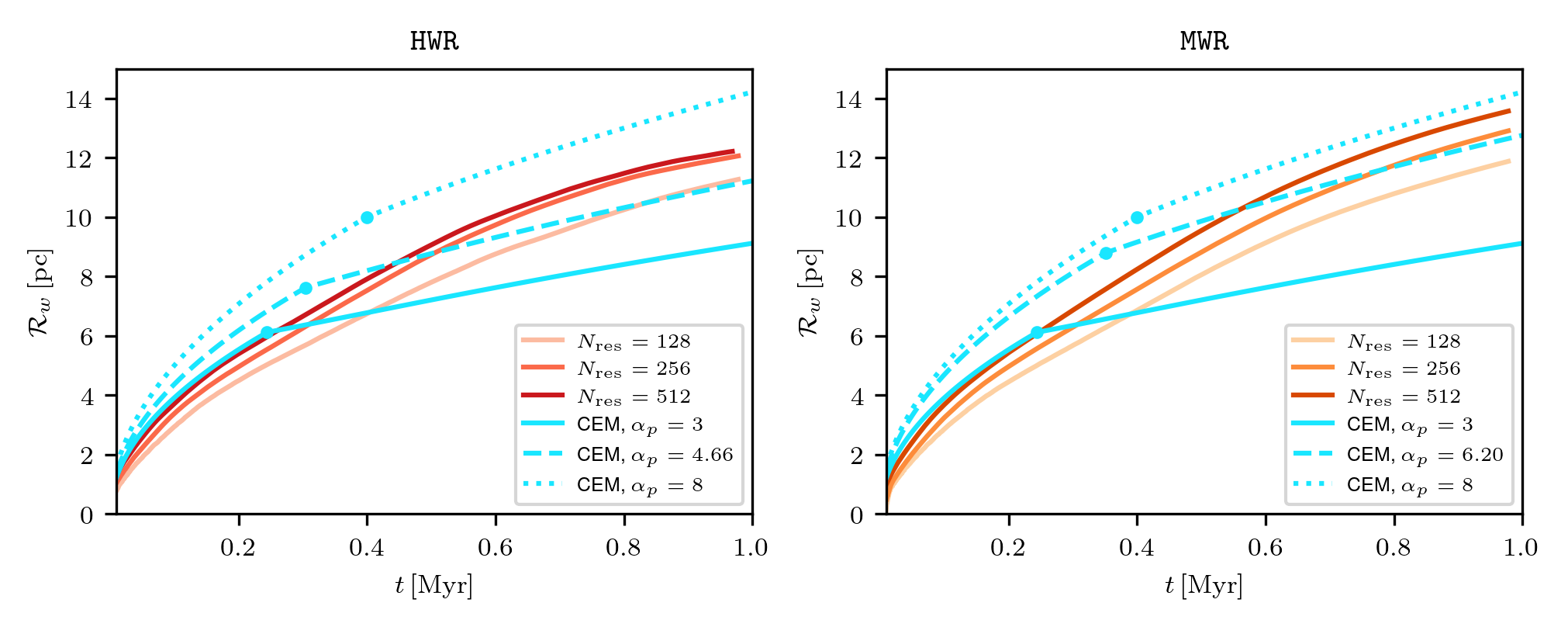}
    \caption{Evolution of the effective wind bubble radius, $\Rw$, in time for the \WRH\, simulations (left, red) and \WRM\, simulations (right, orange). Both panels show comparisons to the CEM theory of \autoref{subsec:theory_joint} in light blue with $\alpha_p = 3$ and $8$ in solid and dotted lines respectively as well as the average $\alpha_p = 4.66,\, 6.20$ values in the \WRH\, and \WRM\, simulations respectively as blue dashed lines. Transition times between the early and co-evolution phases at $\teq$ are marked with blue points on each CEM model curve.}
    \label{fig:wind_radius}
\end{figure*}

\begin{figure*}
    \centering
    \includegraphics{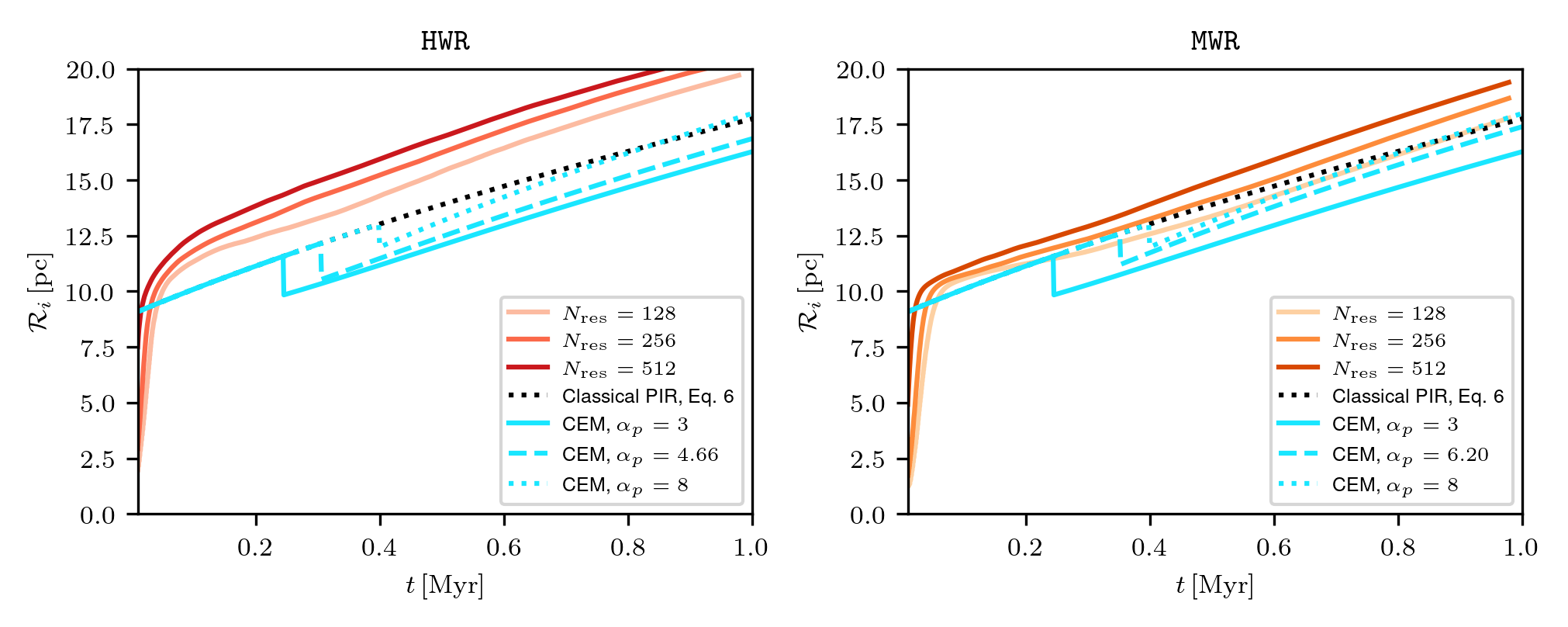}
    \caption{Similar to \autoref{fig:wind_radius} but now showing the volume-effective radius of the ionized gas bubble. The volume used to calculate this radius as $(3V/4\pi)^{1/3}$ includes the volume of the WBB. All CEM models agree at early times $t< \teq$ where they each follow \autoref{eq:spitzer_sol}, shown with a dotted black line in both panels. }
    \label{fig:ionized_radius}
\end{figure*}

\begin{figure*}
    \centering
    \includegraphics{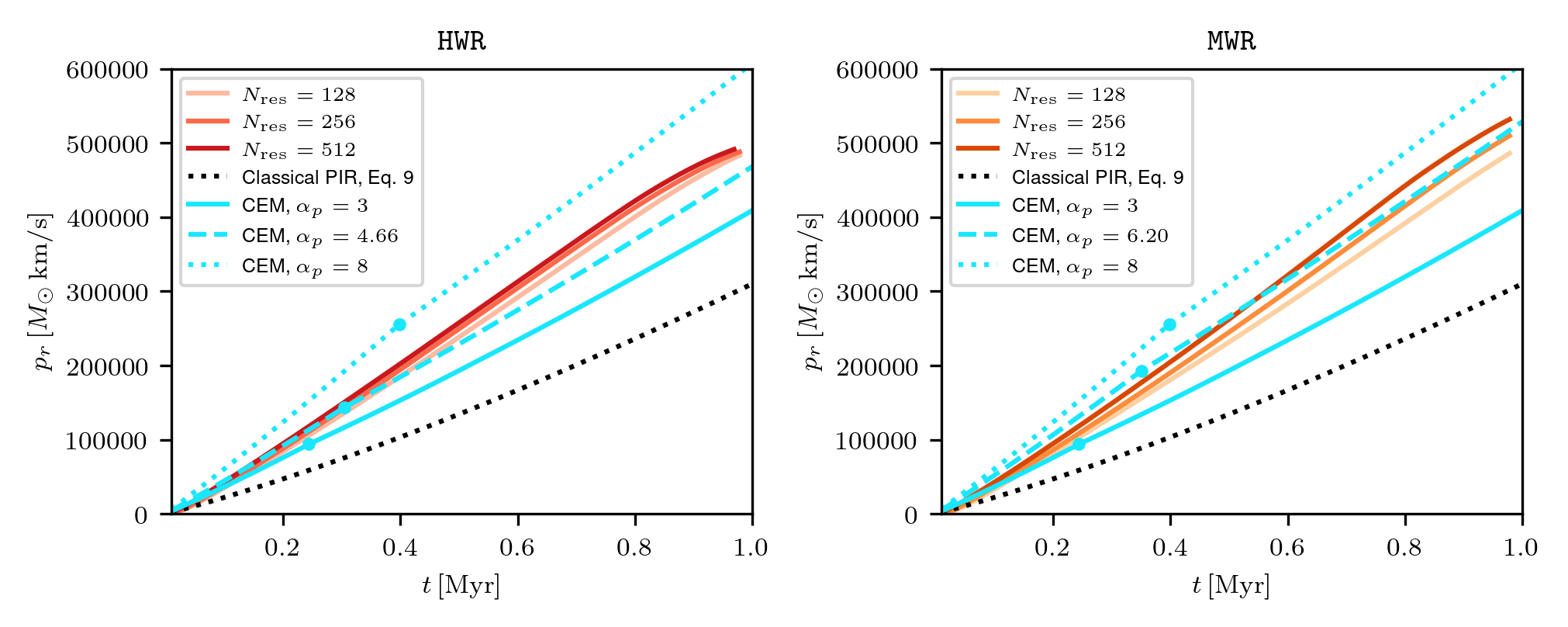}
    \caption{Similar to \autoref{fig:wind_radius} but now showing the total momentum carried by the FBs as measured by comparison to simulations without any feedback (as in the bottom panels of \autoref{fig:momentum_comp}). The same CEM solutions are shown as bright blue lines with the transition between early and co-evolution phases marked with blue dots. The classical momentum evolution given by \autoref{eq:pr_spitzer_adj} is shown as a black dotted line (same as \autoref{fig:momentum_comp}).}
    \label{fig:jfbm_momentum}
\end{figure*}

\subsubsection{Wind-Blown Bubble Expansion}
\label{subsubsec:wind_radius}

In \autoref{fig:wind_radius} we compare the evolution of the volume equivalent radius of the WBBs, $\Rw$, in the \WRH\, and \WRM\, simulations to the predictions from the momentum-driven co-evolution model (MD-CEM) presented in \paperi\ and reviewed in \autoref{subsec:theory_joint}. The theory follows the momentum-driven evolution (\autoref{eq:rEC}) at $t< \teq$ and subsequently is determined by the co-evolution phase of the semi-analytic model.

We see that while both simulations generally follow the $\alpha_p = 3$ CEM well at early times, when the measured instantaneous $\alpha_p$ values are close to this value in the simulations (see left panels of \autoref{fig:wind_evol_comp}), they deviate at later times $t>\teq$ (marked by the dot on the light blue solid lines). In the CEM model the WBB's radial expansion is effectively suppressed by the pressure exerted on the WBB by the PIR, resulting in a somewhat slower expansion. At the same time, $\alpha_p$ also steadily increases in time in the simulations, giving renewed strength to the WBB's expansion. As discussed above, this is mostly caused by the smoothing out of structure in the background, which results in less surface area (at fixed size) over which to cool, resulting in greater retention of energy in the wind bubble and higher $\alpha_p$ (see \autoref{eq:alphap_derive}). This is demonstrated by the greater change in behavior of $\Abub$, compared to $\voutavg$, at $t\lesssim\teq$ in \autoref{fig:wind_evol_comp}. Higher $\alpha_p$ gives renewed strength to the expansion of the WBB causing it to expand above the prediction of the $\alpha_p = 3$ CEM at later times.

While it is reassuring that the $\Rw$ evolution in \autoref{fig:wind_radius} is bounded above by the $\alpha_p =8$ CEM and generally well traced by the average $\alpha_p$ CEM models at late times, the lack of agreement within one model at all times is our first indication of a deficiency of the CEM. Namely, the assumption that $\alpha_p$ is constant is not sufficient to explain the detailed evolution of these bubbles.

We also see a natural tendency to larger $\Rw$ at higher resolution. There are likely two effects at play here. For the \WRM\, simulations, we see a tendency to larger $\alpha_p$ with higher resolution (\autoref{fig:wind_evol_comp}). This increased dynamical impact at higher resolution implies that the WBB is able to expand faster, resulting in a larger WBB volume. However, since $\alpha_p$ changes only mildly with resolution in the \WRH\, simulations, this effect does not explain the $\Rw$ resolution dependence here. In general, as we show in \autoref{app:supp_analysis}, the higher resolution simulations allow for larger clumping factors in their background medium. This leads to the second important effect: large clumping factor means that there are more low-density channels through which the WBB can expand its volume without needing to accelerate a large quantity of mass. This is likely the dominant effect explaining the resolution dependence of $\Rw$ in the \WRH\, simulations. Indeed, though we do not show it explicitly here, we also see $\Rw$ increase with resolution in the \WOH\, simulations even though the effective $\alpha_p$ is decreasing.

\subsubsection{Ionized Bubble Radial Expansion}
\label{subsubsec:jfbm_radius}

In \autoref{fig:ionized_radius} we show the expansion of the volume-equivalent PIR radius, $\Ri$, in the \WRH\, (left, red) and \WRM\, (right, orange) simulations. Predictions of the CEM are shown in light blue. We first reiterate that the discontinuities in the model predictions here are inherent to the choice of an instantaneous transition between the `early evolution phase' ($t<\teq$) and the `joint evolution phase' ($t>\teq$). There are two key effects in the simulations which are not properly accounted for in the CEM that impact the comparison made here.

The first is the inhomogeneity of the background medium. As we discuss further in \autoref{app:spitzer_momentum} and \autoref{app:supp_analysis}, the inhomogeneity in these simulations generally makes the medium more porous to LyC radiation, resulting in a larger PIR than in the uniform density case. As we show through analysis of the clumping factor of the background medium in \autoref{app:supp_analysis}, this effect is somewhat less important in the MHD runs, where the density contrasts are suppressed by the presence of the magnetic field, resulting in a density field that is closer to uniform. The effect of the magnetic field on the porosity of LyC radiation is visually apparent in the right-hand panels of \autoref{fig:mega_plot} in the \WRH\, and \WRM\, simulations, where it is clear that channels emanating out from the central source are ionized to great distances. It is likely due to this effect that $\Ri$ in the simulations is larger than predicted by the CEM models in either set of simulations, but especially the \WRH\, simulations.

This effect is also the cause of larger $\Ri$ values at higher resolution since, as we show in \autoref{app:supp_analysis}, the clumping of the background gas is larger at high resolution. Additionally, as discussed in \autoref{app:spitzer_momentum}, the more faithful representation of the early R-type evolution of the ionization front leads to larger $\Ri$ at higher resolution.

In principle, the presence of the WBB should also shrink the size of the PIR by creating a higher density shell that recombines more rapidly. While this effect is taken into account in the later, co-evolution phase, it is not present at earlier times in the model, where it is assumed that the ionized gas bubble simply follows the Spitzer solution (\autoref{eq:spitzer_sol}, black dotted line in \autoref{fig:ionized_radius}). Since the evolution of $\Ri$ in the simulations is virtually always greater than that predicted by the CEM, this effect is likely subdominant to the increase in ionized gas volume caused by the inhomogeneity. However, since the Spitzer solution is not subject to this effect ($\RSp \geq \mathcal{R}_{i,{\rm CEM}}$ at all but the latest times when the WBB drives the PIR beyond the Spitzer solution) this brings the Spitzer solution into closer agreement with the simulations. This apparent agreement is artificial since neither model takes into account the dominant effect causing large $\Ri$: the inhomogeneity of the background.


\subsubsection{Momentum Evolution}
\label{subsubsec:jfbm_momentum}

In \autoref{fig:jfbm_momentum} we compare the CEM predictions for the total radial momentum carried by the bubble to the \WRH\, and \WRM\, simulations. Here we measure the total momentum with respect to a reference simulation without any feedback, as is used for the bottom panels of \autoref{fig:momentum_comp}. Overall, there is remarkably good agreement between the model and simulations, although, especially in the \WRH\, simulations, the total momentum carried by the bubbles is slightly larger than that predicted by the CEM models. This is due largely to the inhomogeneous background through two distinct effects (discussed also in \autoref{app:spitzer_momentum}). As we saw in \autoref{subsubsec:jfbm_radius}, the ionized gas radii are larger in the inhomogeneous backgrounds due to low-density channels that become ionized. This gives the ionized gas a larger effective surface area over which to exert a pressure force, increasing its dynamical impact. At the same time, inhomogeneities in the background create clumps which are subject to the rocket effect \citep{Bertoldi89,Bertoldi90}. This is most clear in the pressure slice of the \WRH\, simulation in \autoref{fig:mega_plot}, where one can see the tips of high-pressure pillars at the edge of the PIR with lower pressure gas (photo-evaporative flows) just interior to them. This rocket effect is not taken into account in the CEM model and can potentially increase the dynamical impact of the PIR, though this is hard to quantify exactly. The fact that the CEM models are in better agreement with the \WRM\, simulations, which are less inhomogeneous than the \WRH\, simulations, is another indication that these effects are at play in the discrepancies we see.

We also see a moderate increase in $\pr$ with resolution in both the \WRH\, and \WRM\, simulations. In the \WRM\, simulations, this is likely caused by the increased dynamical impact of the WBB at higher resolution. This is indicated by the increase in $\alpha_p$ at higher resolution for the \WRM\, simulations in panel \textit{(d)} of \autoref{fig:momentum_comp}. There is little variation to $\alpha_p$ in the \WRH\, simulations, so the increased momentum in these simulations is likely caused by the more porous, clumpy medium at higher resolution (see \autoref{app:spitzer_momentum} and \autoref{app:supp_analysis}). The clumpier media at higher resolution lead to larger momentum through the same mechanisms discussed above: larger ionized-gas volumes leading to larger ionized-gas pressure forces and, potentially, the rocket effect.

Finally, we note that all simulations lie roughly a factor of 2 above the momentum evolution predicted by \autoref{eq:pr_spitzer_adj} for a PIR evolving in isolation. This emphasizes the dynamical importance of the WBB in the evolution of the FBs simulated here.

\begin{figure*}
    \centering
    \includegraphics[width=\textwidth]{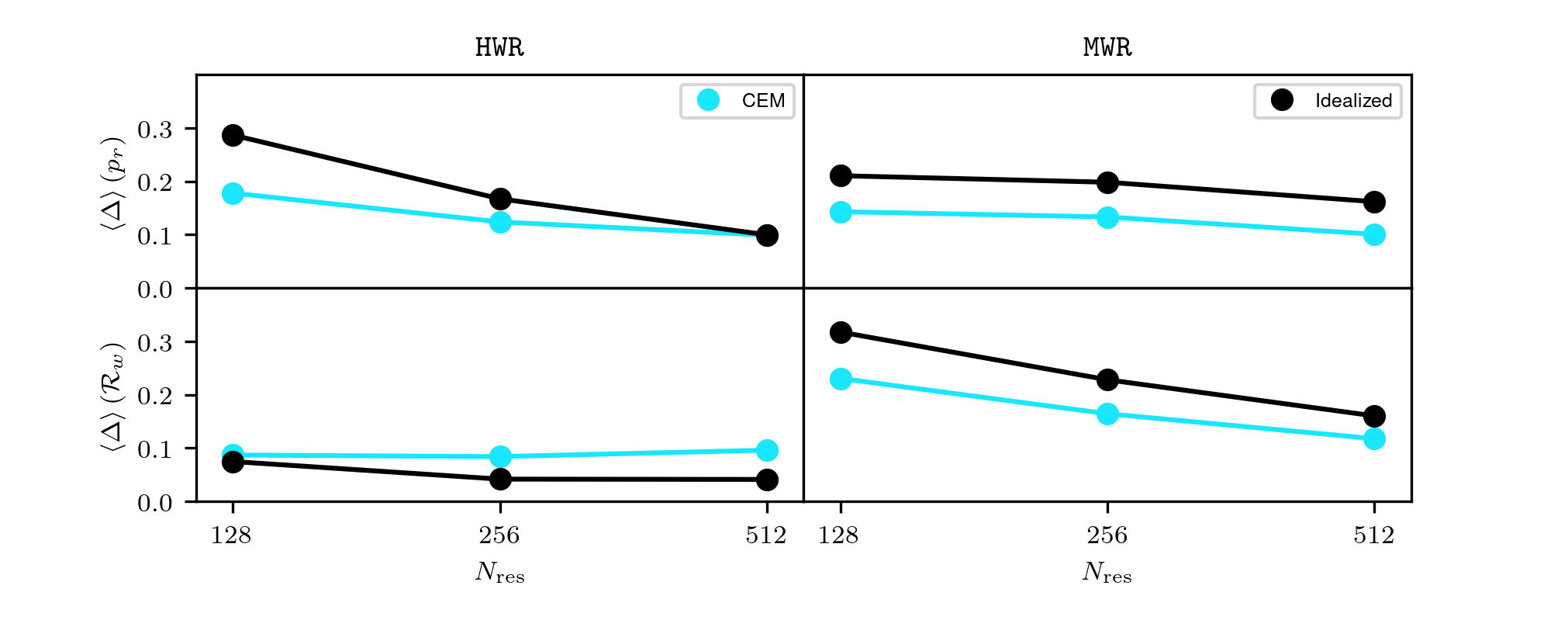}
    \caption{A comparison of the average percent difference between dynamical quantities as measured in the simulations and those predicted by the CEM (light blue) and simplified, uncoupled models (black). Quantities shown are total momentum ($\pr$, top panels) and wind radius ($\Rw$, bottom panels). Points of reference for the black lines are the sum of \autoref{eq:pr_spitzer_adj} and \autoref{eq:pr_EC} for $\pr$ and \autoref{eq:rEC} for $\Rw$. \textit{Left panels}: comparison for the \WRH\, simulations. \textit{Right panels}: comparison for the \WRM\, simulations.}
    \label{fig:cem_comp}
\end{figure*}

\begin{deluxetable*}{cccccccccccccc}
\tablecaption{CEM Comparison.\label{tab:cem_comp}}
\tablewidth{0pt}
\tablehead{
\colhead{Simulation Name} & \colhead{$\left\langle \alpha_p\right\rangle$} & \multicolumn{2}{c}{$\Delta_{\rm CEM}\left(\Rw\right)$} & \multicolumn{2}{c}{$\Delta_{{\rm MD},\alpha} \left(\Rw\right)$} & \multicolumn{2}{c}{$\Delta_{\rm CEM} \left( \Ri\right)$} & \multicolumn{2}{c}{$\Delta_{\rm Sp}\left(\Ri\right)$}  & \multicolumn{2}{c}{$\Delta_{\rm CEM} \left( \pr\right)$} & \multicolumn{2}{c}{$\Delta_{{\rm MD},\alpha +{\rm Sp}} \left(\pr\right)$} \\
& & $\left\langle \Delta \right\rangle$ & $\left\lceil\Delta\right\rceil$ & $\left\langle \Delta \right\rangle$ & $\left\lceil\Delta\right\rceil$ & $\left\langle \Delta \right\rangle$ & $\left\lceil\Delta\right\rceil$ & $\left\langle \Delta \right\rangle$ & $\left\lceil\Delta\right\rceil$ & $\left\langle \Delta \right\rangle$ & $\left\lceil\Delta\right\rceil$ & $\left\langle \Delta \right\rangle$ & $\left\lceil\Delta\right\rceil$}
\startdata
\WRH\texttt{\_N128} & 4.64 & 0.18 & 0.62 & 0.29 & 0.62 & 0.16 & 0.21 & 0.10 & 0.12 & 0.09 & 0.72 & 0.07 & 0.72 \\
\WRH\texttt{\_N256} & 4.78 & 0.12 & 0.40 & 0.17 & 0.40 & 0.20 & 0.26 & 0.14 & 0.15 & 0.08 & 0.53 & 0.04 & 0.53 \\
\WRH\texttt{\_N512} & 4.57 & 0.10 & 0.19 & 0.10 & 0.20 & 0.23 & 0.30 & 0.18 & 0.20 & 0.10 & 0.18 & 0.04 & 0.18 \\
\WRM\texttt{\_N128} & 5.57 & 0.21 & 0.80 & 0.32 & 0.80 & 0.04 & 0.09 & 0.02 & 0.05 & 0.14 & 1.27 & 0.21 & 1.27 \\
\WRM\texttt{\_N256} & 6.20 & 0.20 & 0.63 & 0.23 & 0.63 & 0.07 & 0.13 & 0.03 & 0.07 & 0.13 & 0.93 & 0.20 & 0.93 \\
\WRM\texttt{\_N512} & 6.82 & 0.16 & 0.40 & 0.16 & 0.40 & 0.10 & 0.16 & 0.08 & 0.09 & 0.10 & 0.48 & 0.16 & 0.48
\enddata
\tablecomments{Deviations between simulation measurements for various quantities and models for their evolution in simulations containing WBB and LyC radiation feedback.}
\end{deluxetable*}

\subsubsection{Summary Model Comparison}
\label{subsubsec:model_comp}

In order to provide a quick point of reference for the reader on the effectiveness of our CEM models relative to idealized models of FB properties discussed here we present a quantitative comparison in \autoref{fig:cem_comp} and \autoref{tab:cem_comp}. In particular, for each physical quantity, $Q$, of a FB (which could be $\Rw$, $\Ri$, or $\pr$) we calculate
\begin{equation}
    \label{eq:del_def}
    \Delta_{\rm mod} \equiv \frac{|Q_{\rm sim} - Q_{\rm mod}|}{Q_{\rm sim}} \, ,
\end{equation}
where $Q_{\rm sim}$ is the value taken in the simulation and $Q_{\rm mod}$ is the value predicted by the model (either the CEM or idealized models). For example, $\Delta_{\rm CEM}(\Rw) = |\Rw - \mathcal{R}_{w,{\rm CEM}}|/\Rw$. In order to provide the fairest comparison for the models, we use the average $\alpha_p$ values calculated at each resolution in each simulation; these are given in the second column of \autoref{tab:cem_comp}.

In \autoref{fig:cem_comp} we compare the time-averaged values of $\Delta$ (denoted by the angle brackets) for each of $\pr$ and $\Rw$ in the top and bottom panels respectfully. Light blue lines indicate $\left\langle\Delta\right\rangle$ as compared with predictions of the CEM. Black lines indicate $\left\langle\Delta\right\rangle$ as compared with (i) the sum of \autoref{eq:pr_EC} and \autoref{eq:pr_spitzer_adj} for $\pr$ and (ii) \autoref{eq:rEC} for $\Rw$. In \autoref{tab:cem_comp} we additionally show the maximum percentage difference between the model and the simulations for both the idealized and CEM models as $\left\lceil \Delta \right\rceil$, as well as similar quantities for $\Ri$, which is compared against \autoref{eq:spitzer_sol}.

For the selected FB properties, \autoref{fig:cem_comp} shows that, overall, the predictions of the CEM are within 25\% of the simulated values across resolution. Additionally, the CEM does as well or better than the idealized uncoupled models in each case except modestly for $\Rw$ in the \WRH\, simulations.

If we were to compare the performance of our models in predicting $\Ri$ to the idealized Spitzer model we would see somewhat better performance of the Spitzer model. However, as discussed in \autoref{subsubsec:jfbm_radius}, the idealized Spitzer model for $\Ri$ only appears to do better than the CEM because it predicts larger values for $\Ri$ (not accounting for increased recombination) whereas the main cause for the larger $\Ri$ is the background inhomogeneity, which neither model takes into account. The apparent better agreement in this case is then artificial. For this reason we do not include $\Ri$ in \autoref{fig:cem_comp}.

Additionally, there is no clear pattern towards smaller values of $\left\langle\Delta\right\rangle$ with resolution across different physical parameters and simulations. While the CEM and idealized models clearly improve at higher resolution for $\pr$ in the \WRH\, simulations and $\Rw$ in the \WRM\, simulations these trends are likely due to coincidences. In particular, better agreement at higher resolution for $\Rw$ in the \WRM\, simulations is likely due to the larger $\alpha_p$ values obtained in these simulations. This leads to larger $\Rw$ predictions for both the idealized model and CEM, in better agreement with the large $\Rw$ in the simulations that is likely caused by clumping effects discussed in \autoref{subsubsec:wind_radius}.

Overall, while the CEM does better than the idealized models, it is clear that both evolving $\alpha_p$ and inhomogeneity of the background are as important as the coupling of the feedback mechanisms in predicting the dynamical structure of FBs.

\section{Discussion}
\label{sec:discussion}

\subsection{Review of Past Work}
\label{subsec:past_models}

The question of the evolution of both WBBs and the photoionized region surrounding young, massive stars has been approached from a numerical perspective by a number of studies in the past. In the following section we compare these past efforts to the current work, grouping them based on their approach.

\begin{deluxetable*}{cccccc}
    \tablecaption{Summary of Simulations from the Literature.\label{tab:lit_sim_sum}}
    \tablewidth{0pt}
    \tablehead{
    \colhead{Reference} & \colhead{Dimensionality} &\colhead{Feedback Source} & \colhead{Stellar Evolution}  & \colhead{Time Simulated} & \colhead{Background}  }
    \startdata
    \citet{GSF96} & 1D & $\approx 17\, \Msun$ Star & Constant &  $5\times 10^{-2}$ Myr & U, HD \\
    \citet{vanMarle05} & 1D \& 2D & $40\, \Msun$ Star & MS$\to$RSG$\to$WR &  $\approx 5$ Myr & U, HD \\
    \citet{Dwarkadas22} & " & " & " & " & U, Hydro \\
    \citet{ToalaArthur11} & 1D & $40-60\, \Msun$ Stars & Variable &  $\approx 6$ Myr & U, HD \\
    \citet{Freyer03,Freyer06} & 2D & $35$ \& $60\, \Msun$ Stars & Variable &  $4-5$ Myr & U, HD \\
    \citet{Ngoumou15} & 3D & $\approx 23\, \Msun$ Star & Constant &  $0.5$ Myr & U, HD \\
    \citet{Geen15a} & 3D & $15\, \Msun$ Star & Variable  &  $20$ Myr & U, HD \\
    \citet{Haid18} & 3D & $12$, $23$, \& $60\, \Msun$ Stars & Variable & $2.5$ Myr & U, HD \\
    \citet{Geen21} & 3D & $30$, $60$, \& $120\, \Msun$ Stars & Variable & $\lesssim 1$ Myr & T, MHD \\
    \citet{Geen23} & 3D & $35\, \Msun$ Stars & Variable & $\lesssim 1$ Myr & T, MHD \\
    \citet{Dale14} & 3D & $3,\, 6 \times 10^3\, \Msun$ Cluster & Variable & $6-8$ Myr & T, HD, SG \\
    \citet{Guszejnov22} & 3D & $0.2,\, 1.8,\, 4 \times 10^3\, \Msun$ Cluster & Variable & $6-8$ Myr & T, MHD, SG \\
    \citet{Polak24a} & 3D & $0.3,\, 6.5,\, 85 \times 10^4\, \Msun$ Cluster & Variable & $\approx 1-10$ Myr & T, MHD, SG \\
    This Work & 3D & $5\times 10^3\, \Msun$ Cluster & Constant &  $1$ Myr & T, (M)HD \\
    \enddata
    \tablecomments{All works described here simulate feedback from both photoionizing radiation and stellar winds. Abbreviations on the final column are, uniform (U), turbulent (T), hydrodynamical (HD), magneto-hydrodynamical (MHD), and self-gravitating (SG). \citet{vanMarle05} and \citet{Dwarkadas22} have a three-step, piecewise constant evolution with parameters corresponding to the main sequence (MS), red supergiant (RSG), and Wolf-Rayet (WR) phases. \citet{Dale14} and \citet{Guszejnov22} run further simulations which we do not discuss here.}
\end{deluxetable*}

\begin{figure}
    \centering
    \includegraphics[width=\columnwidth]{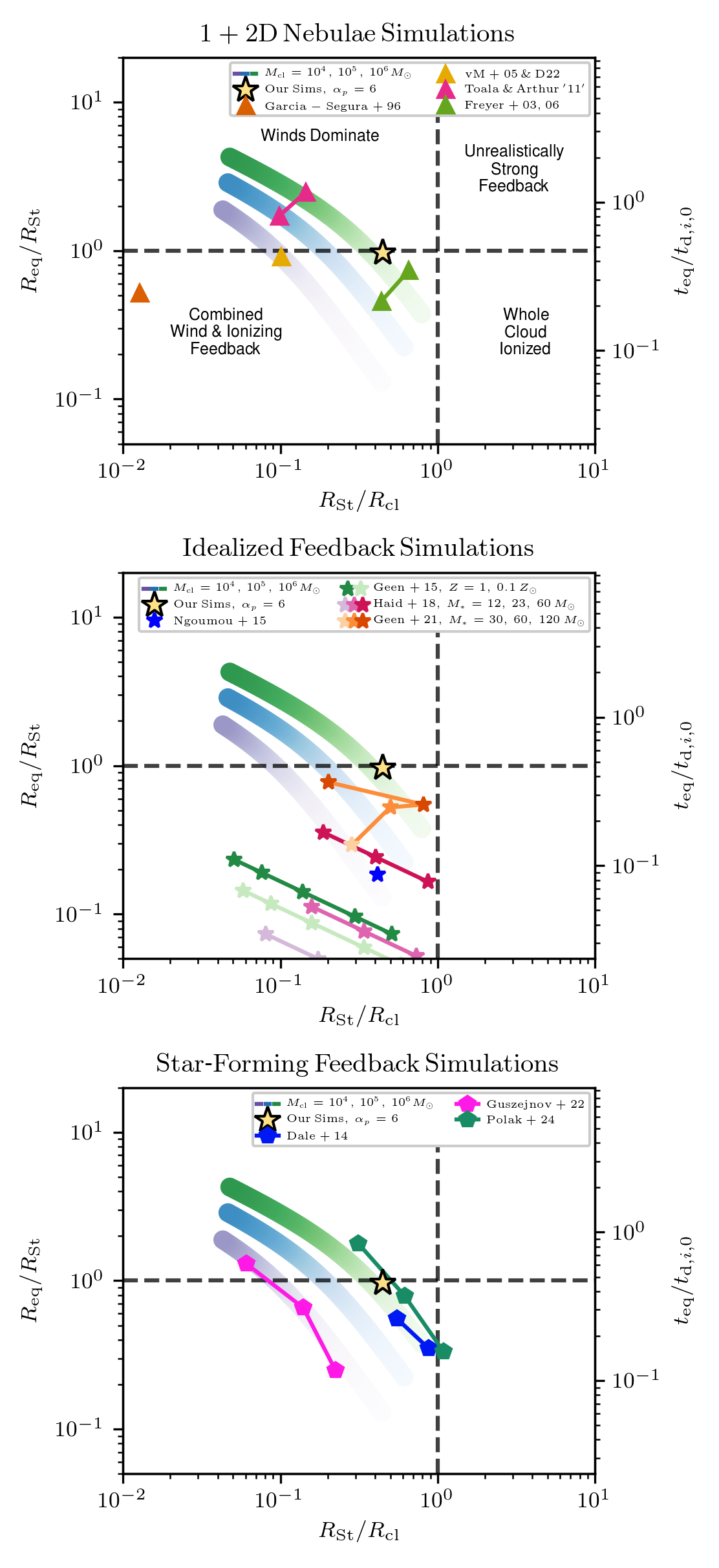}
    \caption{Each panel is identical to Figure 1 of \paperi. Briefly, the y-axis quantifies the importance of WBBs relative to the PIR while the x-axis quantifies the ability of feedback to ionize a large fraction of the clouds. We show where our simulations lie in this space compared to other similar works summarized in \autoref{tab:lit_sim_sum}. \textit{Top Panel}: Simulations run in one and two dimensions. \textit{Middle Panel}: Idealized simulations run in three dimensions. \textit{Bottom Panel}: Star-forming Cloud simulations. Simulations from the same paper with similar parameters are connected with lines in order to guide the eye.}
    \label{fig:feedback_ratio_comp}
\end{figure}

\subsubsection{Idealized Numerical Simulations}
\label{subsubsec:ideal_sim_review}

In addition to the analytic works mentioned above, there have been a number of numerical investigations of the interaction of WBBs and photoionized gas. Here we discuss the subset of these works that are somewhat idealized in that, like the simulations presented here, their star formation is put in `by hand' and the emphasis is on the effect of a FB in a pre-determined environment. In \autoref{fig:feedback_ratio_comp} we provide a comparison of where these simulations lie in the parameter space of wind and photoionization feedback described by \paperi. In particular we show the radius at which the momentum-driven WBB is in pressure equilibrium with the photoionized gas, $\Req$, compared to the \strom radius, $\RSt$, versus $\RSt/\Rcl$. Details on the parameters used for this comparison and where they come from are given in \autoref{app:sim_comp}.

To quickly summarize what follows in detail below, past simulations have largely fallen into two categories. The first are investigations of the structure of circum-stellar gas in massive star nebulae \citep{GSF96,GSML95a,GSML95b,GS96a,GS96b,vanMarle05,Dwarkadas22,ToalaArthur11,Freyer03,Freyer06} which have been entirely performed in one and two dimensional simulations and largely conclude that winds are important for the structure of the circumstellar gas but do not comment at length about the relative dynamics of the WBB versus the photoionized medium. Due to the low dimensionality and uniform backgrounds they simulate, these WBBs are likely not subject to strong cooling due to turbulent mixing at their interfaces \citep{Lancaster21b} and therefore behave in a nearly energy-driven manner, following the \citet{Weaver77} solution. It is then generally clear from where these simulations lie in \autoref{fig:feedback_ratio_comp} (top panel, triangle markers), where we have estimated $\alpha_p$ values based on their energy-conserving like behavior (see \autoref{app:sim_comp}), that the WBBs should be dynamically important in these simulations.

The second category, more comparable to the current work, have focused on the relative dynamical impact of WBBs and the photoionized gas in three-dimensional simulations \citep{Ngoumou15,Geen15a,Haid18,Geen21,Geen23}, but have focused on feedback from individual massive stars (see \autoref{tab:lit_sim_sum}). These simulations have, for the most part, concluded that stellar wind feedback is subdominant to feedback from photoionized gas, which is again in keeping with where they lie in \autoref{fig:feedback_ratio_comp} ($\Req \ll \RSt$, middle panel, star markers).

One key conclusion of our analysis in \paperi\ (also in \citet{Lancaster21c,KrumholzMatzner09}) is that winds are more important when considering feedback from a cluster of stars together than an individual massive star. It is then not surprising that past simulation works have largely concluded that winds are unimportant: they were not simulating environments where one might expect them to be. We provide more specific discussion of each case below.

\citet{GSF96} provided the first such investigation in a one-dimensional hydrodynamical simulation. Their work emphasized the stagnation of FB expansion in the context of ultra-compact HII regions and the presence of dynamical instabilities at the bubble interfaces which they demonstrate with two-dimensional simulations that only include photoionizing radiation (no winds). They do, however, also provide a first investigation of the equilibrium state between WBBs and the photoionized gas (their Equation 16 is analogous to Equation 32 of \paperi). \citet{vanMarle05} perform one and two-dimensional simulations of the evolution of the circum-stellar environment around a $40\,\Msun$ star through the Wolf-Rayet stage to supernova but mainly focus on the structure of the medium rather than the dynamical impact of the feedback mechanisms. They also only simulate the main-sequence evolution of the star (comparable to this study) in one dimension. \citet{Dwarkadas22} performs one and two-dimensional simulations following the same stellar parameter evolution in time as \citet{vanMarle05}. This work also emphasizes the importance of interface instabilities that lead to mixing of the gas phases and cooling at the WBB boundary. These instabilities are enhanced by the inclusion of stellar evolution such as the Wolf-Rayet phase, which can lead to re-acceleration of the WBB front and subsequent Rayleigh-Taylor instabilities. The time-scales for this evolution, however, are longer ($\gtrsim 3\,{\rm Myr}$) than the time-scales probed in our work. Both \citet{vanMarle05} and \citet{Dwarkadas22} use $\Qo$ values during the main-sequence phase that are low for $40\,\Msun$ star, resulting in relatively strong winds. This is reflected both in their simulation outputs and where their simulations lie in \autoref{fig:feedback_ratio_comp}.

\citet{ToalaArthur11} perform one-dimensional simulations of feedback from individual massive stars ranging in mass from $40-60\,\Msun$, focusing mostly on the bookend cases. They also include thermal conduction, which, in keeping with \citet{Weaver77} and \citet{ElBadry19}, helps to cool the WBBs earlier and reduces their dynamical impact. They account for stellar evolution in their simulations and perform two-dimensional simulations for the post-main-sequence evolution of each of the stars they study, focusing on the structure of the inter-stellar material. While they do not extensively comment on the relative dynamical impact of the wind and photoionized gas, it is clear from the relative size of the WBB in their simulations (see their Figure 5) that the WBB has an important dynamical role. This is consistent with where their simulations lie in \autoref{fig:feedback_ratio_comp}. Note, we have assumed $\alpha_p = 75$ for their simulations (see \autoref{app:sim_comp}), which is reflective of their nearly energy-driven behavior. This is at least somewhat an artifact of the 1D simulation which cannot account for turbulent mixing at the WBB's interface.

\citet{Freyer03,Freyer06} perform two-dimensional versions of the FBs from $60$ and $35\, \Msun$ stars considered by \citet{GS96a,GS96b} in one-dimension\footnote{The same picture was considered in a slightly simplified theoretical and numerical context in \citet{GSML95a,GSML95b}.}. Both pairs of works additionally simulate the post-main sequence (PMS) evolution of the stars out to $4$ and $5\, \Myr$ respectively (this evolution was also captured over a smaller domain in two-dimensions in \citet{GS96a,GS96b}). In this PMS the stars enter Luminous Blue Variable (LBV) and Red Supergiant (RSG) phases for the $60$ and $35\, \Msun$ respectively, before becoming Wolf-Rayet stars. The works emphasize the impact on the circum-stellar structure of this evolution. Though they do not comment extensively on the dynamics, it is clear from Figure 18 of \citet{Freyer03} and Figure 17 of \citet{Freyer06} that the dynamical impact of the wind is significant in the $60\,\Msun$ and marginal in the $35\,\Msun$ case, in keeping with the placement of these simulations in \autoref{fig:feedback_ratio_comp}.

Three-dimensional simulations of the joint feedback of winds and ionizing radiation for parameters ($\mdot$, $\Qo$) appropriate for individual stars have been performed by \citet{Ngoumou15}, \citet{Geen15a}, \citet{Haid18}, and \citet{Geen21}.

\citet{Ngoumou15} provides a first investigation of this interaction with a smoothed particle hydrodynamics simulation and assumes (as in \citet{Geen20} and \citet{Dale13a,Dale14} below) that the winds are momentum-driven with $\alpha_p = 1$. They simulate the feedback from an individual star of spectral type O7.5 ($\Mst \approx 23 \, \Msun$ see \citet{Drainebook} Table 15.1) into a uniform background medium and show the the dynamical impact of momentum-driven winds is negligible compared to the impact of ionizing radiation in this scenario. In particular, they compare the evolution of the ionization front radius in their simulation with both ionizing radiation and winds to the model of \citet{Raga12c} (with $\lambda = 0$, which is identical to the \citet{Spitzer78} solution) and find good agreement, indicating the wind has little dynamical impact. As we can see in \autoref{fig:feedback_ratio_comp}, this simulation is clearly in the parameter regime where we wouldn't expect winds to make a dynamical difference ($\Req \ll \RSt$), consistent with their results.

\citet{Geen15a} model feedback from a $15\, \Msun$ star of both solar and tenth solar metallicity in environments with uniform densities ranging from $\nHbar = 0.1-100 \pcc$. They generally find that winds are negligible contributors to the dynamics of the FB. This makes sense in our picture, as we can see from where these simulations lie in \autoref{fig:feedback_ratio_comp}. In fact, we assumed a generous $\alpha_p = 20$ (justified in \autoref{app:sim_comp}) for these simulations and they still lie in the $\Req \ll \RSt$ regime.

\citet{Haid18} simulate three different stellar mass feedback sources in a number of uniform density backgrounds ranging from CNM-like conditions (similar to this work) to conditions in the WNM (more representative of the galactic disk average)\footnote{They also simulate low-density $\nH = 0.1\pcc$ conditions which we do not compare against here, see \autoref{app:sim_comp}.}. They find that winds play a marginally important dynamical role in their most massive star simulations in the densest environments and a more important role in the most massive star simulations in the $\nH = 1\pcc$ WNM-like environment. The former is in keeping with where this simulation lies in \autoref{fig:feedback_ratio_comp} where the $\Req/\RSt = 0.35$ for this case (all other values are smaller). This would be comparable to our work if $\alpha_p = 1$ in our simulations. The WNM-like simulation is slightly inconsistent with our picture, however, this is likely due to the fact that we've estimated the $\alpha_p$ to use in these comparisons based on the dense simulations (as these are the only values readily available from plots in \citet{Haid18}) and the $\alpha_p$ value is higher in the WNM-like conditions. Though it is not possible to directly confirm this from the information available in \citet{Haid18}, this could reasonably be due to a change in the way that numerical diffusion induced cooling acts at the WBB interface as a function of density.

\citet{Geen21} simulate feedback from individual massive stars of $30$, $60$, and $120\,\Msun$ in turbulent, magnetized clouds. These $10^4\, \Msun$ clouds are self-gravitating, magnetized, and set up with an initially spherically symmetric density profile that has a central core and approaches an isothermal sphere ($\rho \propto r^{-2}$) at large radii. The clouds are initialized with some turbulence and allowed to run with no feedback until $120\,\Msun$ worth of mass has accumulated in sink particles, at which point the feedback from the star under study in the particular simulation is `turned-on' from the position of the most massive sink particle. In terms of the \textit{mean} densities, magnetic fields, and turbulence the environment of most of their runs (the \texttt{DIFFUSE} runs) has similar properties to our simulations but for a much lower cloud mass, resulting in a smaller cloud at fixed density. Combined with the density profile of the background, this leads to much more rapid venting of the wind and feedback overall \citep[e.g. Champagne Flows][]{TT79,Franco90}, which could lead to rapid loss of wind energy through bulk-advection of hot gas \citep{HCM09}. Indeed, in \citet{Geen23} similar simulations are run with several different random seeds for the turbulent velocity fields and it is clear that the random seed used in \citet{Geen21} was particularly subject to leakage of the feedback to the cloud exterior (see Figure 12 of \citet{Geen23}). Additionally, comparing the mass of the cloud with the masses of the stars that are simulated for feedback it is worth noting that, with a $\sim 1\%$ SFE as one might expect in these low-mass clouds the chances of having a star of mass greater than or equal to  $30$, $60$, or $120\,\Msun$ is roughly  $25$, $7$, $1\%$ for a Kroupa IMF. These simulations are then fairly interpreted (at least in the $\Mst > 30\,\Msun$ cases) as the effect of these feedback mechanisms on the immediate environment surrounding the massive stars rather than the entire cloud in which they form. We provide this information as explanation for why the simulations of \citet{Geen21} are likely not the best test environment for the dynamical interaction of photoionizing radiation and WBBs, even if they are quite interesting for the study of champagne flows and specific observational comparisons to the Orion cloud \citep{Geen17}. With that in mind, \citet{Geen21} conclude that, in their set of simulations, stellar wind feedback is broadly unimportant in adding significant momentum to the surrounding cloud: contributing at most 10\% of the resulting momentum. This is broadly consistent with where these simulations lie in \autoref{fig:feedback_ratio_comp}, even though this analysis does not take into account the steep density gradient in the material surrounding the feedback source. In particular, comparing the placement of the $120\, \Msun$ star run in the \texttt{DIFFUSE} cloud set up in \autoref{fig:feedback_ratio_comp} (dark orange star at large $\RSt/\Rcl$) with Figure 1 of \citet{Geen21}, we correctly predict that nearly the entire cloud should be ionized by the star. The fact that \citet{Geen21} find little wind impact for the  $120\, \Msun$ star run in the \texttt{DENSE} cloud set up is likely due to quick venting of the WBB, as we mentioned above.

\citet{Geen23} simulate the same environmental set-up as the \texttt{DIFFUSE} clouds from \citet{Geen21} but for a $35\,\Msun$ star and running various simulations with different random seeds for the turbulent background velocities as well as a control suite for different implementations of relevant physics. Feedback parameters for these simulations are not directly provided in the work so we do not plot these simulations in \autoref{fig:feedback_ratio_comp}, though they are likely comparable to the $30\,\Msun$ simulations of \citet{Geen21}. \citet{Geen23} emphasize the role of WBBs in reducing the effects of photoionizing radiation by trapping the radiation in the shell of the WBB. Given that the density profiles in these simulations are close to that of an isothermal sphere, this finding is in keeping with our analysis in Appendix A of \paperi\ as well as their work in \citet{Geen22}. As we discuss in \paperi, we expect this effect to be important around the cores of individual massive stars, but not on the scale of GMCs overall.


\subsubsection{Star-Forming Cloud Simulations}

Though simulations that form stars self-consistently under the action of gravity along with both stellar wind and photoionization feedback are beyond the scope of the current work, we will briefly review here works that have performed such simulations. Broadly, these simulations fall into two categories as well. The first employ momentum-driven or mass-loaded stellar wind feedback for the sake of computational cost \citep{Dale13a,Dale14,Lewis23,CCC23,Polak24a,Ali22} which inherently decreases the potential effectiveness of WBB feedback. The second inject the full mechanical energy of winds, with the potential for them to be energy-conserving if they don't succumb to efficient cooling. The only simulations in this second category, to our knowledge, that include the effects of both photoionizing radiation and winds are those of \citet{Guszejnov22} using the framework of \citet{STARFORGE21} (though simulations by the current authors have accounted for these effects separately \citet{JGK18,JGK19,JGK21,Lancaster21c}). We show where the simulations of \citet{Dale14}, \citet{Guszejnov22}, and \citet{Polak24a} lie in the space of wind versus photoionizing feedback in the final panel of \autoref{fig:feedback_ratio_comp}. These simulations span the regions of parameter space from where we might expect winds to be unimportant ($\ReqRSt\ll 1$) to where they should be dynamically dominant ($\ReqRSt >1$). \citet{Dale14} provide a direct comparison of the effects of WBB and PIR feedback and conclude that winds are not dynamically dominant, consistent with the location of these simulations in \autoref{fig:feedback_ratio_comp}. \citet{Guszejnov22} provide a comparative analysis of the effects of different feedback mechanisms on the IMF (the main object of their study) and final SFE for the cloud parameters represented by the middle pink pentagon in \autoref{fig:feedback_ratio_comp}. For this case they conclude that winds are unimportant for regulating star formation (see their Figure 3). This is somewhat in tension with the placement of this simulation in \autoref{fig:feedback_ratio_comp} ($\ReqRSt \approx 0.7$), which suggests that winds should play a role in the dynamics. However, this is likely a consequence of the relatively small clouds simulated in \citet{Guszejnov22}, which allow only a handful of massive stars to form. In this case our use of IMF averaged quantities to make \autoref{fig:feedback_ratio_comp} would over-estimate the impact of WBBs.

Stellar winds and ionizing radiation have also been included in simulations on the scale of galaxies \citep{Andersson24,Gatto17,Peters17,Rathjen21} or even in cosmological contexts \citep{Calura15,Calura22,Calura24,Lahen23}. However, given the resolution requirements to properly model the dynamics of WBBs and photoionized gas \citep{Pittard21,Pittard22Rad,Deng24,Lancaster24a} it is reasonable to question the faithful representation of these feedback processes in some of these environments.

\subsection{Summary of the Roles of Turbulence}
\label{subsec:role_of_turbulence}

Here, we summarize the effects of the turbulent backgrounds that we simulate on the impact of the stellar feedback mechanisms. The primary role of turbulence in the background media that we simulate is to create density inhomogeneity. As was shown in \citet{Lancaster21b} and \citet{Lancaster24a}, the primary effect of this inhomogeneity on the evolution of WBBs is to seed turbulent mixing at the WBB interface which has the potential to drastically reduce the dynamical impact of winds through mixing-enabled cooling at the WBB interface. Difficulties related to truly resolving mixing at these interfaces is discussed further in \autoref{subsec:problems}.

As is discussed in detail in \autoref{app:spitzer_momentum}, \autoref{app:supp_analysis}, and \autoref{subsec:joint_sim_comp}, inhomogeneity/clumping in the background results in larger $\Ri$ and $\pr$. This correlation is seen across resolution (higher resolution, more inhomogeneous) and in comparing HD (more inhomogeneous) to MHD (less inhomogeneous) simulations. Larger ionized gas volumes ($\Ri$) are likely created in more clumpy media as a larger degree of clumping leads to more low-density channels through which ionizing radiation can propagate to large distances, ionizing a large fraction of gas. This may have an impact on the momentum imparted to the surroundings as well, as the larger size for an ionization front over which the PIM exerts a pressure force results in more momentum. The increased dynamical impact due to larger ionized gas volumes is likely a subdominant effect as (i) these volumes are preferentially lower density, and therefore lower pressure (as they are nearly isothermal at the ionized gas temperature) (ii) these ionization fronts tend to be short lived as dense clumps of gas move transversely, occulting parts of the ionized gas from their ionizing source. The primary source of the increased momentum in the clumpier media is likely the so-called ``rocket-effect'' \citep{Bertoldi89,Bertoldi90}, which results in strong evaporation of gas at D-critical ionization fronts with dense gas, though it is hard to quantify this exactly.

As the presence of turbulence in the background medium generally decreases the dynamical impact of WBBs and increases the dynamical impact of LyC radiation, in the simplest sense turbulence should increase the relative importance of LyC radiation compared to WBBs. However, as is the main motivating point of this investigation, no effect occurs in isolation in star-forming regions. In particular, as we discuss in \autoref{subsec:PIR_cooling_effect} and \autoref{subapp:shell_thick}, the presence of the PIM increases the dynamical impact of WBBs through reducing the degree of cooling at their interfaces.

We do not expect the results of this work to depend strongly on the specific realization of the turbulent background as long as the background is statistically isotropic and homogeneous on scales compared to the FB radius. Though changes in the power spectrum of turbulence used to generate the background field may result in different properties of the inhomogeneities, we expect the general trends with clumping of the background medium will still be followed.

\subsection{Prospects for Future Work}
\label{subsec:problems}

There are several processes relevant to feedback from massive stars that we have not yet included. Though beyond the intended scope of this work, the most obvious of these are additional feedback mechanisms like direct radiation pressure on gas and dust grains \citep{Draine11,Raskutti16,raskutti17,JGK16,JGK18}, indirect radiation pressure on dust \citep{KrumholzMatzner09,Fall10,Murray10,Skinner15,Menon22,Nebrin24} or even proto-stellar jets \citep{STARFORGE21,Cunningham11,Cunningham18,NakamuraLi07}, though the last are only expected to be relevant for the formation of individual stars \citep{MatznerMcKee00}. We additionally do not include the creation or transport of cosmic rays in the star-forming region, though this likely has little dynamical impact on the FB \citep{Murray10}.

As noted in \autoref{sec:simulations}, the choices made for the parameters of the simulations presented here were meant to be representative of FBs driven by a cluster of stars within a GMC. As discussed in Appendix A of \paperi, and is evident in much of the literature on individual stars discussed in \autoref{subsec:past_models}, the evolution of FBs around individual massive stars may evolve very differently, particularly in regards to radiation trapping \citep[e.g.][]{Geen22,Geen23}. In \paperi\, we argue that the percolation of FBs from individual massive stars, and their joining together into a cluster-driven FB, should happen relatively quickly compared to the star formation times of GMCs, motivating our focus on the dynamics of cluster-driven FBs. However, there are certainly many observational studies of feedback around individual massive stars \citep{SOFIA_FEEDBACK,Bonne22} whose quantitative comparison to the results of this work should be performed cautiously.

As discussed in \citet{Lancaster24a}, the key open problem in understanding WBB dynamics is a comprehensive picture of the interplay of mixing, cooling, and thermal conduction at their interfaces, particularly in the presence of magnetic fields. While we do not include thermal conduction in this work, the results of \citet{Lancaster24a} show that the scales relevant to resolve these diffusive processes are often far below the resolving power of global simulations \citep[see][Figure 2]{Lancaster24a} such as those run here or elsewhere in the literature. This means that, even if we were to include thermal conduction in our evolution equations, the resolution we can afford would not accurately represent its impact on the gas structure or FB dynamics. The solution to accurately representing the combination of conduction and turbulent mixing/cooling physics in such simulations is the development of sub-grid models calibrated against smaller scale, resolved simulations \citep{WeinbergerHernquist23,arkenstone1,arkenstone2,Butsky24}, but this has yet to be carried out for this context. If we are able to understand this physics we would also be able to include them in a semi-analytical model for the HII region evolution. This would allow for models that do not assume purely momentum-driven (fixed $\alpha_p$) or energy-driven \citep{Weaver77}-like behavior for the WBB but account for the energy-loss dynamics on-the-fly.

As evidenced by the comparisons made in \autoref{sec:results}, more advanced semi-analytical models must account for inhomogeneities in the background media into which they expand if they want to reproduce key features of FBs. This is currently not accounted for in any non-numerical model and could have important implications for observable aspects of HII regions \citep[e.g.][]{BalserWenger24,MD24,Nemer24}.

\section{Conclusion}
\label{sec:conclusion}

We end here with a short summary of our main conclusions:
\begin{itemize}
    \item We perform three-dimensional MHD simulations (described in \autoref{sec:simulations}) of HII region evolution in the part of the star-formation parameter space where we expect both winds and photoionized gas to play important dynamical roles. In \autoref{subsec:past_models} we provide a thorough comparison to previous work and show that this work is the first FB simulation run in this part of parameter space, where past studies had focused on feedback from individual stars. We vary the feedback mechanisms included and use these as tests to understand the effects of each mechanism.

    \item In \autoref{subsec:PIR_cooling_effect} we analyze the effect that the presence of the photoionized gas has on the dynamics of the WBB in the context of the analysis of \citet{Lancaster24a}. We conclude that the photoionized gas increases the dynamical impact of WBBs in two ways. First, it decreases the ability for WBBs to cool at their interfaces by removing collisional Ly$\alpha$ excitation as a cooling mechanism. Second, similar to the effect of magnetic fields in \citet{Lancaster24a}, photoionized gas acts to smooth density inhomogeneities in the background medium which decreases the surface area over which the WBBs can cool (as shown in panels \textit{(c)} and \textit{(f)} of \autoref{fig:wind_evol_comp}). Both of these effects lead to higher energy retention in the WBB in our simulations, which therefore gives higher pressure and greater dynamical impact for WBBs.

    \item In \autoref{subsec:joint_sim_comp} we compare the semi-analytical model of \paperi\ (reviewed in \autoref{sec:theory}) to our simulations and find agreement to within 20\% for $\Rw$ and $\pr$, and 30\% for $\Ri$ (\autoref{fig:cem_comp}). We conclude that the major sources of disagreement are due to the model's lack of accounting for variable momentum enhancement factors, $\alpha_p$ (as are present in the simulations, \autoref{fig:wind_evol_comp}) and the inhomogeneities of the cloud material that the FBs expand into. This second effect allows LyC radiation to escape to large radii, resulting in much larger photoionized gas volumes than the uniform medium case and therefore a large discrepancy between the CEM predictions and simulations in $\Ri$ as well as $\Rw$ to a lesser degree. Both variable $\alpha_p$ and background inhomogeneities should be considered in future implementations of semi-analytic models for feedback in star-forming regions.
\end{itemize}

\acknowledgments

The authors would like to thank Ulrich Steinwandel, Laura Sommovigo, Brent Tan, Drummond Fielding, Romain Teyssier, Mike Grudi\'{c}. The authors thank the anonymous referee for their thorough and thoughtful review which improved the quality of this work.
L.L. gratefully acknowledges the support of the Simons Foundation under grant 965367. This work was supported in part by grant 510940 from the Simons Foundation to E.~C.\ Ostriker. C.-G.K. acknowledges support from a NASA ATP award No. 80NSSC22K0717. J.-G.K acknowledges support from the EACOA Fellowship awarded by the East Asia Core Observatories Association, and from KIAS Individual Grant QP098701 at Korea Institute for Advanced Study. GLB acknowledges support from the NSF (AST-2108470, AST-2307419), NASA TCAN award 80NSSC21K1053, and the Simons Foundation through the Learning the Universe Collaboration. 

\software{
{\tt scipy} \citep{scipy},
{\tt numpy} \citep{harrisNumpy2020}, 
{\tt matplotlib} \citep{matplotlib_hunter07},
{\tt adstex} (\url{https://github.com/yymao/adstex})
}

\appendix

\begin{figure*}
    \centering
    \includegraphics{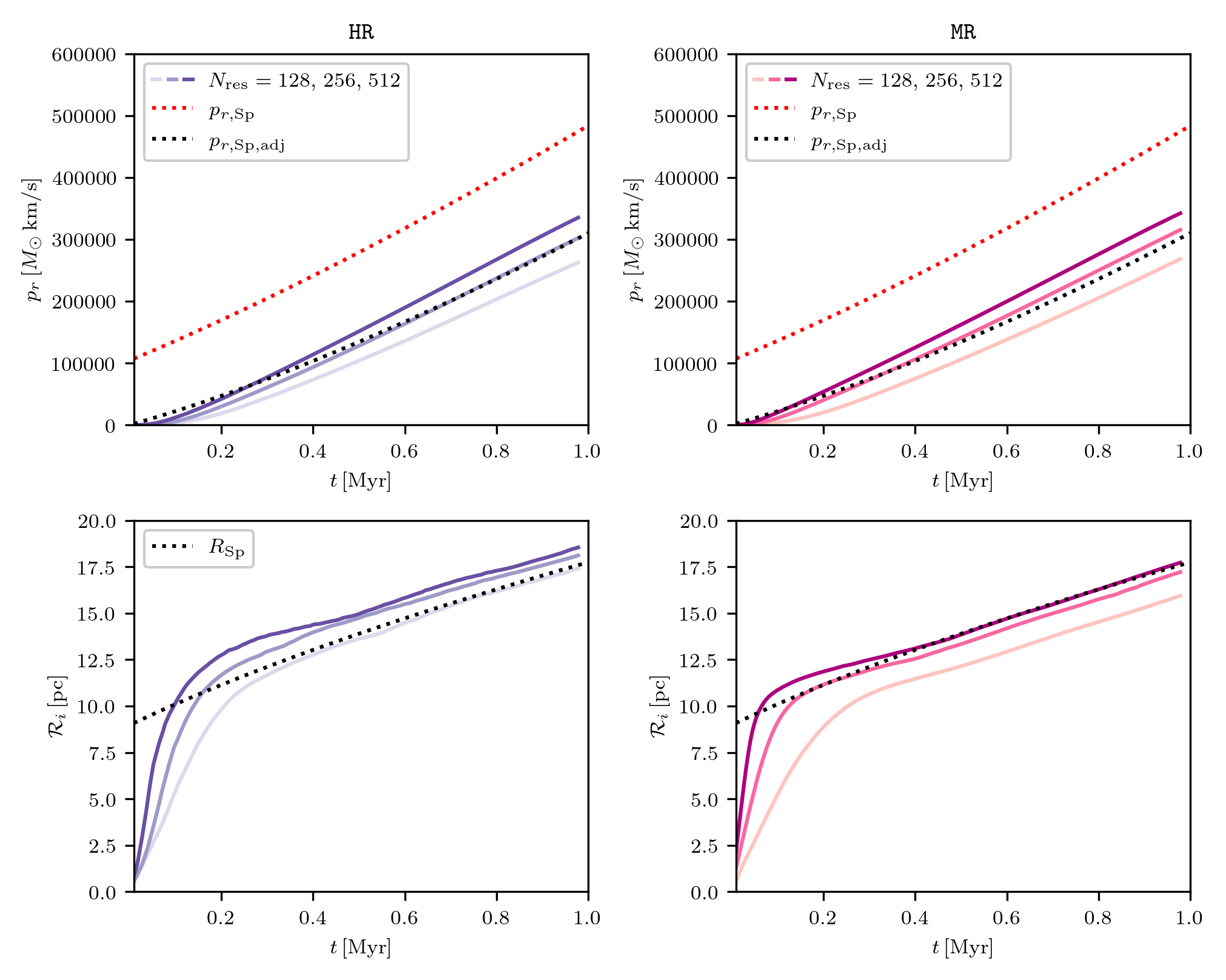}
    \caption{Evolution of the total radial momentum carried by the photoionized gas and shocked neutral gas (top) and its volume equivalent radius (bottom) for the \texttt{HR} (left) and \texttt{MR} (right) simulations at each resolution. For the momentum, we compare the evolution with the predicted radial momentum evolution for the Spitzer solution, as given by \autoref{eq:pr_spitzer1} (red dotted line) and the adjusted Spitzer solution as given by  \autoref{eq:pr_spitzer_adj} (black dotted line). The adjusted solution is clearly a better match to the momentum evolution of the simulated photoionized gas bubbles. For the radial evolution in the bottom panels we show the predicted Spitzer radial evolution as given in \autoref{eq:spitzer_sol}.}
    \label{fig:sptiz_momentum_comp}
\end{figure*}

\section{Validity of Spitzer Momentum Equation}
\label{app:spitzer_momentum}

As discussed in \paperi\ and reviewed in \autoref{sec:theory}, the Spitzer solution, strictly interpreted, implies that the there is a non-zero amount of momentum carried by the PIR at $t=0$, as we can observe from \autoref{eq:pr_spitzer1}. In that section we suggested an adjustment to this formal solution that recovers zero radial momentum at $t=0$ by subtracting the mass interior to the PIR from the mass assumed to be carrying all of the momentum in the bubble's shell. This resulted in \autoref{eq:pr_spitzer_adj}.

In the top panels of \autoref{fig:sptiz_momentum_comp} we compare \autoref{eq:pr_spitzer1} and \autoref{eq:pr_spitzer_adj} to the total radial momentum in our simulations with radiation feedback only. It is clear that the adjusted solution, indicated by the black dotted line, is a much better match to the simulated solution than the strictly interpreted Spitzer momentum, shown as a red dotted line.

We also see clear evidence for differences in the total radial momentum carried by the PIR bubble as a function of numerical resolution, with more momentum being injected at higher resolution. This is likely due to the fact that, at higher resolution the inhomogeneities in the background are more faithfully represented on smaller and smaller scale. This has two effects. Firstly, this allows LyC radiation to take advantage of more `holes' in the density field so that the ionized gas makes up a larger volume at a fixed time. This is illustrated by the evolution of the volume-equivalent radius in the bottom panels of \autoref{fig:sptiz_momentum_comp}. Secondly, the higher resolution allows smaller, denser structures which are more subject to the rocket effect \citep{Bertoldi89,Bertoldi90}.

The fact that $\Ri$ is larger at earlier times in the higher resolution simulations is a numerical artifact due to the fact that we are solving the time-independent radiative transfer equation. In particular, as we choose a time-step for iterating the ray-tracing based on the time-step constraint in the warm ($T<2\times 10^4\Kel$) gas the ray-tracing time is roughly
\begin{equation}
    \Delta t_{\rm rt} \approx \frac{\Delta x}{c_{s,{\rm warm}}} \approx 4 \times 10^3\yr\,\left(\frac{\Nres}{512} \right)^{-1} \, .
\end{equation}
This is long compared to the recombination time $\trec \approx 10^3\yr$ at all resolutions. Due to this choice, the $R$-type evolution of the ionization front is not properly captured.

\section{Supplementary Analysis}
\label{app:supp_analysis}

We present several pieces of supporting data to the arguments presented in the main body of the text.

\begin{figure*}
    \centering
    \includegraphics{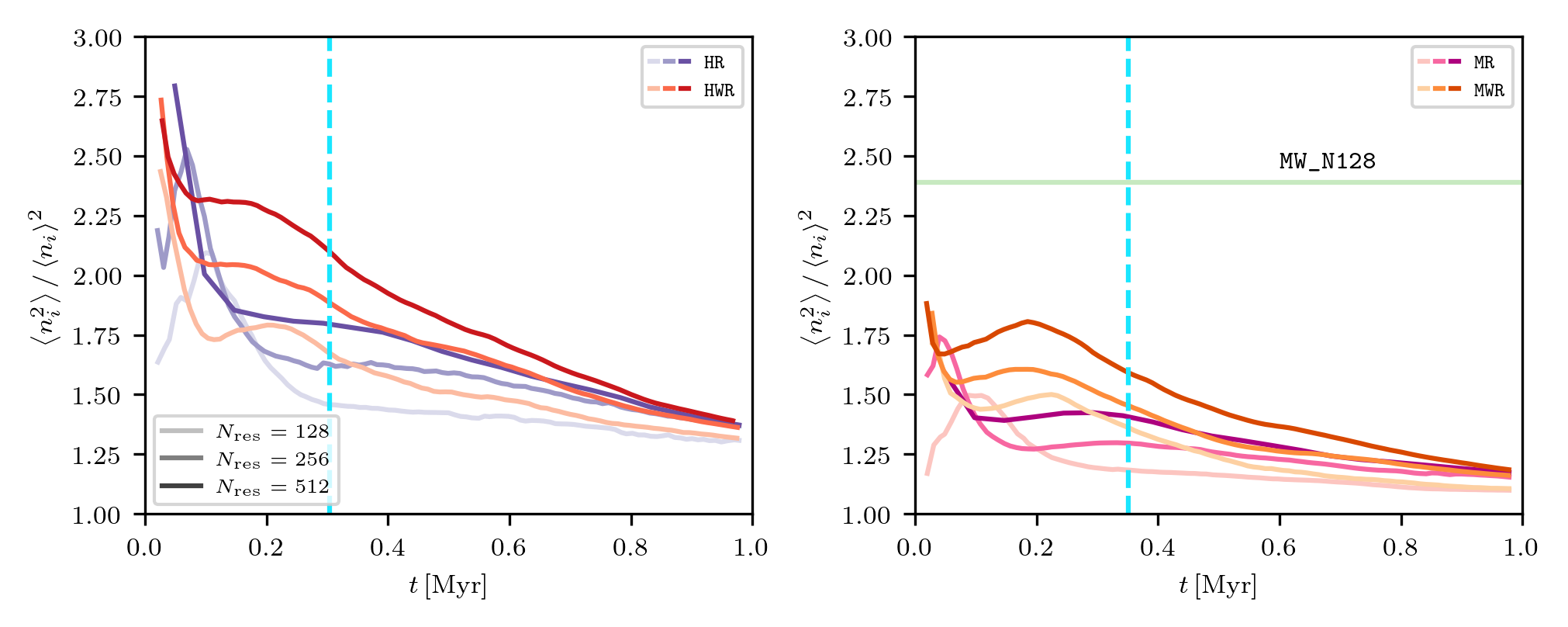}
    \caption{A comparison of the evolution in time of the clumping factor in photoionized gas, $\clumpi$, across different simulations. \textit{Left panel}: Comparison of \texttt{HR} and \WRH\, simulations. \textit{Right panel}: Comparison of \texttt{MR} and \WRM\, simulations. The light green line in the right panel indicates the clumping factor of the background medium in the $\Nres = 128$ \WOM\, simulation. The background clumping factors of all other simulations lie above the upper limit of this plot: $\clump = 5.42,\, 7.70,\,\&\, 10.3$ in the $\Nres = 128,\, 256,\,\&\, 512$ \WOH\, simulations respectively and $\clump = 2.39,\, 3.10,\,\&\, 4.24$ in the $\Nres = 128,\, 256,\,\&\, 512$ \WOM\, simulations respectively. Darker lines indicate higher resolution in both panels. Vertical, dashed, light blue lines in both panels indicate $\teq$ for the CEM with $\alpha_p$ given by the time-average values for \WRH\, (left) and \WRM\, (right) models. These are the same values used in \autoref{subsec:joint_sim_comp}.}
    \label{fig:clump_comp}
\end{figure*}


\subsection{Clumping in the Ionized Gas}
\label{subapp:clumping}

In the \autoref{sec:results} we appeal to differences in the clumping structure of the background medium across simulations at different resolutions, and including different physics, to explain several effects. In order to justify these claims, we provide here an analysis of the clumping factors of the medium into which both the PIR and WBB expand. We define the clumping factor as
\begin{equation}
    \label{eq:clump}
    \mathfrak{C} \equiv \frac{\left\langle n^2\right\rangle}{\left\langle n\right\rangle^2}
\end{equation}
where $n$ is the number density of the gas and angle brackets here indicate volume-weighted averages.

We calculate $\mathfrak{C}$ in the simulations both with and without magnetic fields for the background medium after the turbulent evolution described in \autoref{subsec:cloud_details} but before the feedback has been turned on. For the HD simulations we find $\mathfrak{C} = 5.42,\, 7.70,\,\&\, 10.3$ in the $\Nres = 128,\, 256,\,\&\, 512$ simulations respectively. For the MHD we find $\mathfrak{C} = 2.39,\, 3.10,\,\&\, 4.24$ in the $\Nres = 128,\, 256,\,\&\, 512$ simulations respectively. This demonstrates that simulations are clumpier at both higher resolution and in the HD simulations when compared with the MHD simulations. The former effect is likely due to higher resolutions being able to resolve higher density structure, as gas is able to be effectively concentrated in a smaller volume. The latter effect, smaller clumping in MHD simulations, is likely due to the field's pressure support against compression to high density in shocks as is supported by turbulent collapse models for star formation \citep{PN11,HC11}.

In \autoref{app:spitzer_momentum} and \autoref{subsubsec:jfbm_radius} we appeal to the increased clumping of the background medium at higher resolution to explain larger volumes of ionized gas. This is due to more efficient escape of LyC radiation to large radii through lower density channels. Out analysis above clearly shows that this is plausible based on measurements of clumping in the background that the LyC radiation expands in to.

In \autoref{subsec:momentum} we appeal to smoothing out of structure in the photoionized gas by sound waves in order to help explain the reduction in the surface area of the WBB, $\Abub$, in simulations with LyC radiation. To demonstrate that this effect is taking place, in \autoref{fig:clump_comp} we show the evolution of the clumping factor in the photoionized medium, defined as in \autoref{tab:phase_defs}. In order to compare the evolution of $\clumpi \equiv \left\langle n_i^2\right\rangle/\left\langle n_i\right\rangle^2$ in simulations with and without stellar winds, we show this on a scale of $1-3$. This precludes comparison to the clumping factors in the backgrounds of the simulations, which are all larger than this range, except for the lowest resolution \WRM\, simulation, whose clumping factor is shown in the right panel. Given that all measured $\clumpi$ are much lower than those in the background, and the WBBs nearly entirely expand into the photoionized gas, the assertion in the \autoref{subsec:momentum} that this decrease in clumping helps to decrease the surface area of the WBBs is reasonable. The assertion that this decrease in clumping is due to sound waves is supported by the near exponential decrease in $\clumpi$ with time in the simulations without wind bubbles (shown in purple at the left of \autoref{fig:clump_comp} and magenta on the right). To be clear, these sound waves still suppress clumping in simulations with WBBs, the decrease is simply not exponential as there are other factors which lead to increased clumping, as we discussed below.

We can also take this opportunity to explore several interesting aspects of the evolution of $\clumpi$ in \autoref{fig:clump_comp}. First, we see in the simulations without stellar wind feedback that the clumping initially increases in time, though only at moderate and low resolution ($\Nres = 128,\, 256$). This is a numerical artifact that, similar to the radial evolution of these bubbles discussed in \autoref{app:spitzer_momentum}, is due to the fact that we are solving the time-independent radiation transfer equations. Due to this, the expansion of the front is delayed (we are not capturing the R-type evolution) and occurs more quickly along lower-density lines of sight, biasing $\clumpi$. As the full \strom sphere is ionized, a maximum in $\clumpi$ is reached and it subsequently declines due to pressure inhomogeneities caused by the density inhomogeneities and relatively constant pressure. This effect is less prominent at higher resolution since the R-type ionization front evolution is more faithfully captured when the time step is shorter.

The time step is also much shorter in the simulations containing stellar wind feedback, as the presence of hotter gas decreases the simulation time-step. This is also the explanation for the more rapid decrease in $\clumpi$ in these simulations compared to those without winds. We also see that, after a rapid decrease, $\clumpi$ in the simulations with winds then levels out or even increases before decreasing again. This is due to the enhancement in density by the shell of the WBB in the photoionized gas (discussed further below). That is, as predicted by the CEM model, the WBB compresses the ionized gas in its shell, causing a larger range in densities and an increase in $\clumpi$. This conclusion is supported by the fact that the value of $\clumpi$ begins to decrease again right around $\teq$, indicated by the vertical, dashed blue lines in \autoref{fig:clump_comp}. It is at this time, $\teq$, that the WBB should effectively be compressing all of the PIR equally so that uneven compression should no longer cause an increase in $\clumpi$.

\begin{figure*}
    \centering
    \includegraphics{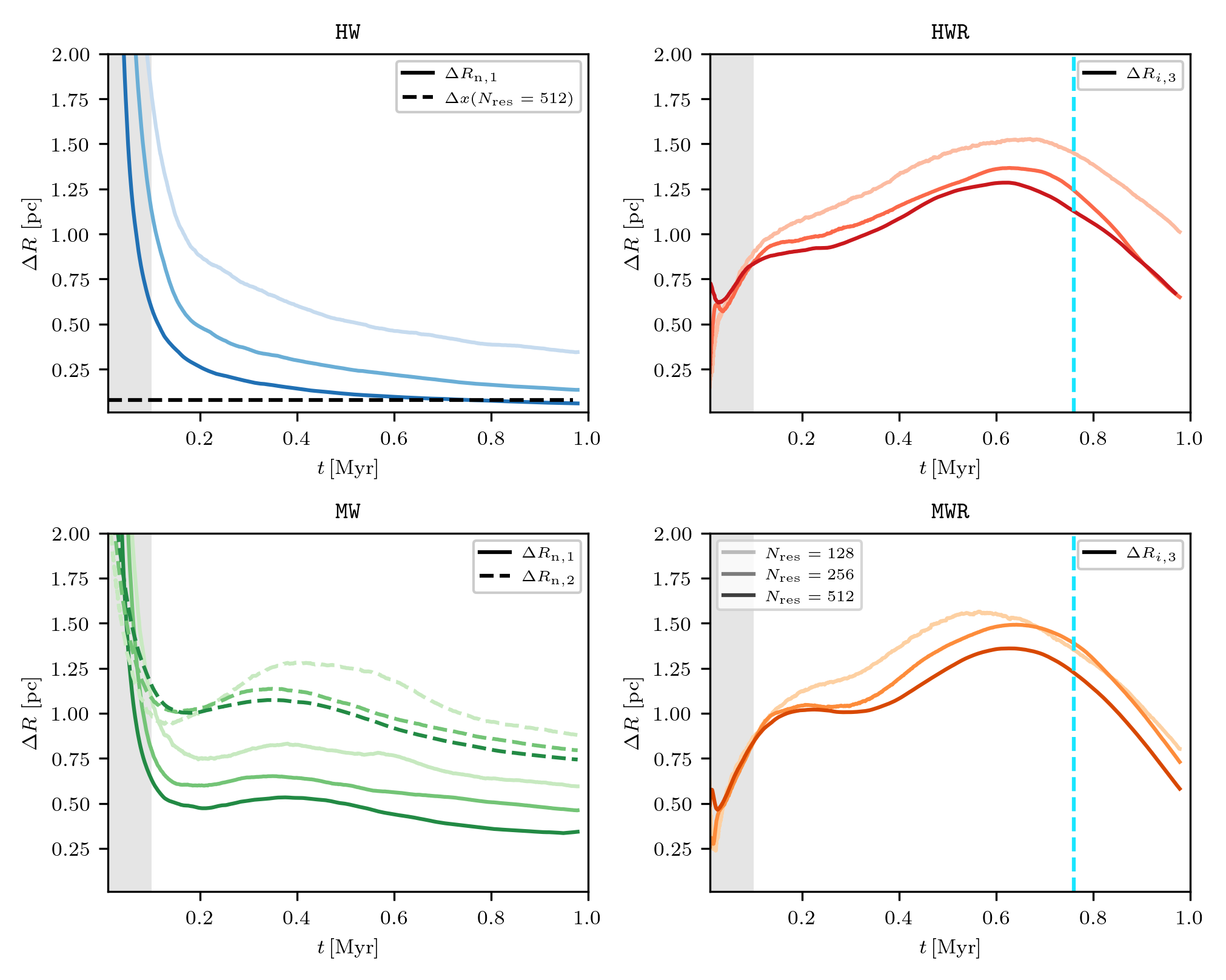}
    \caption{Estimates for the evolution of the thickness of the shell around the WBB, $\Delta R$. Clockwise from the top-left panels indicate estimates for the \WOH, \WRH, \WRM, and \WOM\ simulations. Each panel uses different methods for calculating $\Delta R$, as detailed in \autoref{subapp:shell_thick}, with two different methods compared in the bottom left panel. The darker lines in each panel correspond to higher resolution simulations. The dashed black line in the top left panel indicates the resolution in the highest resolution simulations. The grey areas in each panel roughly indicate time periods where the shell thickness estimates are less trustworthy.}
    \label{fig:shell_thick}
\end{figure*}

\subsection{Thickness of the WBB Shell}
\label{subapp:shell_thick}

In \autoref{subsec:PIR_cooling_effect} and \citet{Lancaster24a} we appealed to the presence of a `thick'-shell surrounding the WBB in MHD simulations or those including LyC radiation feedback to explain the decrease in $\Abub$ through the suppression of so-called ``thin-shell instabilities'' \citep{Vishniac83}. Though the presence of such a shell is apparent in the higher pressure and momentum density region surrounding the WBB in \autoref{fig:mega_plot}, it is nonetheless useful to quantify the degree to which the thickness of the shell changes between the simulations presented here, in order to verify this interpretation. In \autoref{fig:shell_thick} we attempt to quantify this behavior.

In a simulation with a uniform background medium it is straightforward to select the volume of gas that consists of the shell as it has higher density and non-zero velocity \citep{KO15,IH15,JGK_NCR23}. In our turbulent simulations, it is much harder to make a simple distinction between the shell and the surrounding gas, as the background itself has large variations in density and velocity \citep{IH15,KimOstrikerRaileanu17}. In the \WOH\, simulations the shell of the WBB is well-traced by wind-polluted gas which diffuses across the contact discontinuity. This is demonstrated by the agreement between the momentum measured in wind-polluted material and that measured in comparison to a reference simulation for \WOH\, simulations in \autoref{fig:momentum_comp}. However, as discussed in \autoref{subsec:momentum}, momentum is clearly moved out of wind-polluted material in the \WOM, \WRH, and \WRM\, simulations. This is our first indication of the presence of a thick-shell, though it precludes using $\fwind$ to trace the shell material.

In order to try to quantify the thickness of the WBB we try several different methods of selecting for the volume of the shell. All methods make selections based on a stress quantity we define as
\begin{equation}
    \label{eq:stressq}
    \mathfrak{s} \equiv \rho v_r^2 + \frac{\gamma}{\gamma-1} P + \frac{B^2}{2}\, ,
\end{equation}
that is, the sum of the radial Reynolds stress, the enthalpy, and the magnetic energy density of the gas (written in units such that the magnetic permeability is 1).

We associate gas with the neutral shell that surrounds the WBB in the \WOH\, and \WOM\, simulations by selecting all gas in the WNM, UNM, and CNM (as defined in \autoref{tab:phase_defs}) with $\mathfrak{s}\geq \rhobar \ci^2$, with $\rhobar$ the background average density and $\ci= 10\kms$, the rough ionized gas sound speed. As the background turbulent velocity in the simulations is $v_t \approx 7\kms$, the background Alfv\'{e}n speed in the \WOM\, and \WRM\, simulations is roughly $v_A \approx 2 \kms$, and the background sound speed is $c_s\approx 1 \kms$, this selection works reasonably well, but also tends to select denser regions of gas in the background. This over-selection causes an over-estimate in the volume of the shell. We will refer to the volume of gas selected in this way as $V_{\rm sh, n}$.

We measure the thickness of the shell in neutral gas in two ways. The first is
\begin{equation}
    \label{eq:dRn1}
    \Delta R_{{\rm n},1} \equiv \frac{V_{\rm sh, n}}{\Abub}\, ,
\end{equation}
the volume of the neutral shell divided by the surface area of the bubble. This method assumes that the shell is thin and evenly spread in thickness over the surface of the bubble. It should be most accurate for the \WOH\ simulations, where the shell should indeed be thin. This method is used for the solid lines in the left hand panels of \autoref{fig:shell_thick}.

The second method is defined assuming a spherical geometry as
\begin{equation}
    \label{eq:dRn2}
    \Delta R_{{\rm n},2} \equiv \left(\frac{3}{4\pi}(V_w + V_{\rm sh,n}) \right)^{1/3} - \Rw \, .
\end{equation}
That is, we calculate the volume equivalent radius of the wind and shell volume (the first term above) and subtract out $\Rw$. This should work in simulations which are closer to a spherical geometry, such as the smoother, less fractal \WOM\, simulations, but will likely over-estimate the region size for fractal geometries. This method is used for the dashed lines in the bottom left panels of \autoref{fig:shell_thick}.

In the simulations with LyC radiation we select the shell of ionized gas around the WBB by selecting gas in the PIM (as defined in \autoref{tab:phase_defs}) and requiring either $\mathfrak{s}\geq \rhobar \ci^2$ or $\mathfrak{s}\geq 4 \rhobar\ci^2$. We will refer to the volume of gas selected using the first condition as $V_{{\rm sh},i,1}$ and the volume selected using the second condition as $V_{{\rm sh}, i ,2}$. The first method is more efficient at selecting the shell at later times, when the rest of the ionized gas bubble has had some time to expand so that the ionized gas density is less then $\rhobar$. The second method is more useful at early times, when nearly the whole of the ionized gas has $\mathfrak{s} \geq \rhobar \ci^2$, so one needs a stricter constraint to select the shell. Similar to \autoref{eq:dRn2} above, we use these volumes to define two thicknesses for the shell of ionized gas surrounding WBBs in the \WRH\, and \WRM\, simulations as
\begin{equation}
    \label{eq:dRi12}
    \Delta R_{i,1} \equiv \left(\frac{3}{4\pi}(V_w + V_{{\rm sh},i,1}) \right)^{1/3} - \Rw \,\,\,\, , \,\,\,\,
    \Delta R_{i,2} \equiv \left(\frac{3}{4\pi}(V_w + V_{{\rm sh},i,2}) \right)^{1/3} - \Rw \, .
\end{equation}

As the first method works best at late times and the first method works best at early times we define our third method as
\begin{equation}
    \label{eq:dRi3}
    \Delta R_{i,3} \equiv \left( 1- e^{-\frac{t}{\tdio}}\right)\Delta R_{i,1} + e^{-\frac{t}{\tdio}}\Delta R_{i,2} \, .
\end{equation}
That is, we switch between the two estimates with a time-scale given by the dynamical expansion time of the \strom sphere, $\tdio$, indicated by the vertical, dashed light blue lines in the right hand panels of \autoref{fig:shell_thick}. It is on this time scale that the ionized bubble should begin to expand, the density should decrease, and $\Delta R_{i,1}$ should become a better estimate of the volume. This method is used to calculate the lines shown in the right hand panels of \autoref{fig:shell_thick}.

We have chosen the methods that we use to calculate the thickness of the shell in each of the simulations shown in \autoref{fig:shell_thick} based on which method we expect to be the most reliable. We use $\Delta R_{{\rm n},1}$ for the \WOH\, simulations in the top left as these are quite fractal and have very thin shells. This is apparent from the fact that the thickness of the shell approaches the resolution. Indeed, $\Delta R$ in the \WOH\, simulations are basically identical to one another within factors of 2 corresponding to changes in the resolution. This indicates that the main factor determining the thickness of the shell in the \WOH\, simulations is the resolution. We have included both $\Delta R_{{\rm n},1}$ and $\Delta R_{{\rm n},2}$ in the bottom left panel of \autoref{fig:shell_thick} because the shell is neither very thin (appropriate for $\Delta R_{{\rm n},1}$) nor perfectly spherical (appropriate for $\Delta R_{{\rm n},2}$), so while neither is quite appropriate, these present two limiting cases that the correct answer may be expected to lie between.

We see that estimates of $\Delta R$ in the \WOH\, and \WOM\, simulations diverge to large values at early times. This is simply a sign that estimators for $V_{\rm sh,n}$ are imperfect and include larger fraction of the background volume at earier times when this volume has not been incorporated into the WBB yet. We may, therefore not necessarily trust the values of $\Delta R$ at $t\lesssim 0.1\Myr$ in these simulations.

At late times in the \WRH\, and \WRM\, simulations our estimates of $\Delta R$ drop to small values. This is a somewhat similar problem of not capturing the volume of gas that we would like to based on our stress condition. Specifically, at this point the PIM and the WBB have likely expanded enough that none of the material in the PIM satisfies the conditions on $\mathfrak{s}$ that we have used to define the ionized shell.

Keeping all of these effects in mind, it is clear from the above analysis that the thickness of the shells in the \WOM, \WRH, and \WRM\ simulations are all larger than those in the \WOH\, simulations. In particular, the thickness of the shell seems to be dominated by the sound speed of the PIM in the \WRM\, simulations (rather than the added magnetic stress) as the thickness of the shell in these simulations is comparable to that in the \WRH\, simulations and slightly larger than the \WOM\, runs.

\begin{deluxetable}{cccccccc}
    \tablecaption{Feedback Parameters Used for Comparison.\label{tab:feedback_params_lit}}
    \tablewidth{0pt}
    \tablehead{
    \colhead{Ref.} & \colhead{Feedback Source} & \colhead{$\mdot\, [M_{\odot}{\rm /Myr}]$} & \colhead{$\Lwind\, [{\rm erg/s}]$} &
    \colhead{$\vw\, [{\rm km/s}]$} & \colhead{$\Qo\, [{\rm s}^{-1}]$}
    &  \colhead{Densities $[{\rm cm}^{-3}]$}  & \colhead{$\Rcl\, [{\rm pc}]$}}
    \startdata
    (1) & $\approx 17\, \Msun$ Star & $2.5\times 10^{-2}$ & $3.13\times 10^{34}$ & $2000$ & $10^{48}$ & $10^7$ & $0.05$  \\
    (2) & $40\, \Msun$ Star & $9.1\times 10^{-1}$ & $2.27\times 10^{35}$ & $890$ &  $4.62 \times 10^{47}$ & $13.5$ &  $40$ \\
    (3) & $40\, \Msun$ Star & $5.6\times 10^{-1}$ & $1.34\times 10^{36}$ & $2750$ & $10^{49}$ & $100$ & $30$ \\
    (3) & $60\, \Msun$ Star & $2.37$ & $6.28\times 10^{36}$ & $2900$ & $3.16\times 10^{49}$ & $100$ &  $30$ \\
    (4) & $35\, \Msun$ Star & $3.0\times 10^{-1}$ & $1.5\times 10^{36}$ & $4000$ & $1.0\times 10^{49}$ & 20 & $20$ \\
    (4) & $60\, \Msun$ Star & $1.5$ & $10^{37}$ & $4600$ & $3.5\times 10^{49}$ & 20 & $20$ \\
    (5) & $\approx 23\, \Msun$ Star & $1$ & $1.3\times 10^{36}$ & $2000$ & $10^{49}$ & 30 & $16$  \\
    (6) & $15\, \Msun$ Star, $Z = 0.1Z_{\odot}$ & $5.0\times 10^{-3}$ & $2\times 10^{33}$ & $1126$ & $5.2\times10^{47}$ & $0.1$, $0.5$, $5$, $30$, $100$ &  \\
    (6) & $15\, \Msun$ Star & $10^{-2}$ & $4\times 10^{33}$ & $1126$ & $3.5\times10^{47}$ & $0.1$, $0.5$, $5$, $30$, $100$ &  \\
    (7) & $12\, \Msun$ Star & $2\times 10^{-3}$ & $5\times 10^{33}$ & $2816$ & $1.4\times 10^{48}$ & $1$, $10$, $100$ & $88,\, 41,\, 19$ \\
    (7) & $23\, \Msun$ Star & $6\times 10^{-2}$ & $1.5\times 10^{35}$ & $2816$ & $10^{49}$ & $1$, $10$, $100$ & $88,\, 41,\, 19$ \\
    (7) & $60\, \Msun$ Star & $2$ & $8\times 10^{36}$ & $3563$ & $1.7\times 10^{49}$ & $1$, $10$, $100$ & $88,\, 41,\, 19$ \\
    (8) & $30\, \Msun$ Star & $2.4\times10^{-1}$ & $5\times 10^{35}$ & $2582$ & $6.5\times 10^{48}$ & $106$ & $8.6$ \\
    (8) & $60\, \Msun$ Star & $3.2$ & $6.5\times 10^{36}$ & $2550$ & $3.5\times 10^{49}$ & $106$ & $8.6$ \\
    (8) & $120\, \Msun$ Star & $12.7$ & $3\times 10^{37}$ & $2740$ & $1.5\times 10^{50}$ & $106$ & $8.6$ \\
    (8) & "  & " & " & " & " & $6814$ & $2.2$\\
    (9) & $35\, \Msun$ Star  & ? & ? & ? & ? & $106$ & $8.6$\\
    This Work & $5\times 10^3\, \Msun$ Cluster & $14.8$ & $4.9\times 10^{37}$ & $3230$ & $2\times 10^{50}$ & $86.25$ & $20$  \\
    \enddata
    \tablecomments{These parameters are estimated based on information available in the various cited works. Unless otherwise noted, these parameters were calculated for solar metallicity stars. References: (1) \citet{GSF96}, (2) \citet{vanMarle05} \& \citet{Dwarkadas22}, (3) \citet{ToalaArthur11}, (4) \citet{GS96a,GS96b,Freyer03,Freyer06}, (5) \citet{Ngoumou15}, (6) \citet{Geen15a}, (7) \citet{Haid18}, (8) \citet{Geen21}, (9) \citet{Geen23}. Details for self-gravitating runs are given below.}
\end{deluxetable}

\section{Simulation Literature Comparison}
\label{app:sim_comp}

Here we explain the feedback values we used for our comparison to other simulations given in \autoref{subsubsec:ideal_sim_review} and \autoref{fig:feedback_ratio_comp}. The work of \citet{GSF96} performs a one-dimensional feedback simulation meant to be representative of the dense core around a proto-star. The $\Qo$ value used in this work is appropriate for a $\approx 17\,\Msun$ O9V star \citep{Drainebook}. We choose $\Rcl$ value appropriate for the ultra-compact HII regions they are interested in (see their Figures 2-4).

\citet{vanMarle05} and \citet{Dwarkadas22} use identical feedback parameters, which they claim are meant to be representative of a $40\, \Msun$ star. However, we note that the value of $\Qo$ employed in these simulations is about a factor of $\approx 35$ too small for the mass of star they claim to simulate \citep{Drainebook}. These simulations are carried out in a uniform background density field with $\rhobar = 10^{-22.5}\, {\rm g}\pcc$, which we use to derive the background number density of $\nHbar = 13.5\pcc$. We use $\alpha_p = 10$ which is based on a rough estimate of the ratio of the bubble radius to the free wind radius ($\Rw/\rfree$) in \citet{Dwarkadas22} combined with Equation A19 of \citet{Lancaster24a} (which could be derived from the pressure relation for a momentum-driven like bubble, Equation 7 of \paperi).

For \citet{ToalaArthur11} we approximate the feedback parameters for their $40$ and $60\, \Msun$ stars based on their Figures (1-3), the background density is given in the text. We take $\alpha_p = 75$, estimated using the same method as above. We take an $\Rcl$ value based approximately on the maximum extent of the HII regions in their simulations at the end of the main-sequence phase.

For the simulation parameters considered both by \citet{Freyer03,Freyer06} and \citet{GS96a,GS96b} we approximate the feedback parameters based on Figure 2 of \citet{Freyer03} and Figure 1 of \citet{Freyer06}. We take the LyC luminosities as $L_{\rm LyC} = 3,\, 10\times 10^{38}\, {\rm erg/s}$ which, assuming an average ionizing photon energy of $18\, {\rm eV}$, results in ionizing photon emission rates of $\Qo = 1.0,\, 3.5 \times 10^{49}\,{\rm s}^{-1}$. Background density is given in the text and we take the cloud radius based on the approximate extent of their FBs at the end of the main sequence evolution in their simulations. We take $\alpha_p = 12$ using the same method as above.

For \citet{Ngoumou15} all parameters are given in the text of their paper and are meant to be representative of a spectral type O7.5 star \citep{Smith06}, which has a mass of roughly $23 \, \Msun$ \citep{Drainebook}. We assume a momentum enhancement factor of $\alpha_p = 1$, which is appropriate for their simulations as they assume momentum-driven wind bubbles.

For \citet{Geen15a} we approximate values for the feedback parameters of the solar and tenth solar metallicity, $15\, \Msun$ stars they aim to simulate based off of approximate average values in their Figure 1. a luminosity in the Lyman Continuum ($h\nu > 13.6\, {\rm eV}$) of $L_{\rm LyC} = 1.5, \, 1 \times 10^{37}\, {\rm erg/s}$ which we convert to an ionizing photon rate using an average ionizing photon energy of $18\, {\rm eV}$. We use their simulated densities to derive cloud radii by assuming the cloud has a mass $\Mcl = 10^5 \,\Msun$ resulting in cloud radii of $\Rcl = 190,\, 111,\, 52,\, 28,\, 19.0\, {\rm pc}$. We take $\alpha_p = 20$ using the same logic as above.

We use Figure 1 from \citet{Haid18} to approximate the feedback parameters for each of their simulated stars, taking values based approximately on the first $\Myr$ of evolution. \citet{Haid18} simulate a range of uniform density backgrounds from $\nH = 0.1-100\pcc$ in 1 dex increments. We omit the $0.1\pcc$ simulations from comparison here since (as is pointed out in \citet{Haid18}) for this case the background is ionized by the galactic background radiation and the LyC radiation of the source has very little dynamical effect. We take $\alpha_p = 10,\, 3,\, \& \, 1$ for the $\Mst = 12,\, 23, \&\, 60 \Msun$ cases respectively based approximately on the momentum evolution compared to the injected momentum given in the bottom panels of their Figure 6. These values are for their highest density ($\nH = 100\pcc$) ``Cold Neutral Medium'' simulations and could vary based on density. We derive $\Rcl$ base on density in the same way as for \citet{Geen15a}.

The feedback parameters for \citet{Geen21} are given in the appendix and in particular in Figure A1. We derive approximate values for $\mdot$, $\Lwind$, and $\Qo$ from the first $1\,\Myr$ of evolution in these plots. We derive the wind velocities from $\Lwind$ and $\mdot$. $\Rcl$ is derived from the densities as above. We use $\alpha_p = 4,\, 3\, 2,\, 1$ for the $30\,\Msun$ (DIFFUSE), $60\,\Msun$ (DIFFUSE), $120\,\Msun$ (DIFFUSE), and , $120\,\Msun$ (DENSE) runs respectively, these are estimated based on their Figure 7. \citet{Geen23} references \citet{Geen21} for feedback parameters on the $35\,\Msun$ star simulated in \citet{Geen23}, but these parameters are not provided in \citet{Geen21}.

For the self-gravitating cloud simulations of \citet{Dale14}, \citet{Guszejnov22}, and \citet{Polak24a} we take cloud parameters ($\Mcl$, $\Rcl$, $\sfe$) from their papers where available and assume the same IMF averaged feedback parameters as we use in this work, described in \autoref{sec:simulations}. We additionally assume $\alpha_p = 1$, which is appropriate for \citet{Dale14} and \citet{Polak24a} as these simulations use momentum-driven \citep{Dale14} and significantly mass-loaded \citep{Polak24a} wind prescriptions. This may, however, underestimate the impact of winds in the \citet{Guszejnov22} simulations, although these clouds are quite small and may easily leak wind material, which also leads to momentum-driven-like behavior. For \citet{Dale14} we only show results for their Runs I and UQ as these are the runs we are able to estimate $\sfe$ of based on their Figure 11. We then use $\Mcl = 10^4\,\Msun$ and $\Rcl = 10,\, 5\, {\rm pc}$, $\sfe = 30,\,60\%$ for runs I and UQ respectively. For \citet{Guszejnov22} we show results for their series of simulations at varying surface density and estimate final $\sfe$ from their Figure 8. We take $\Mcl = 2\times 10^4$ and $\Rcl = 10,\, 3,\, 1\, {\rm pc}$ with $\sfe  = 1,\, 9,\, 20\%$ respectively. All the information we need from \citet{Polak24a} is provided in their abstract: $\Mcl = 10^4,\, 10^5,\, 10^6\,\Msun$, $\Rcl = 11.7\,{\rm pc}$, and $\sfe = 36,\, 65,\, 85\%$. 

%

\bibliography{bibliography}{}
\bibliographystyle{aasjournal}

\end{document}